\documentclass[12pt,letter]{article} 
\usepackage{url}
\usepackage{caption}
\usepackage{authblk}
\usepackage{booktabs}
\usepackage[utf8]{inputenc}
\usepackage[margin=1in]{geometry}
\usepackage[table]{xcolor}
\usepackage{amsmath}
\usepackage{longtable}
\usepackage{lscape}
\usepackage[
backend=biber, natbib,
citestyle = authoryear,
style = authoryear,
uniquelist=false,
uniquename = false,
sorting = nyt,
maxbibnames=99,
maxcitenames=2,
dashed=false
]{biblatex}
\usepackage[english]{babel}        
\usepackage{csquotes}             
\usepackage{tikz}
\usetikzlibrary{shapes.geometric, arrows}
\usepackage[titletoc,toc,title]{appendix}
\usepackage{enumitem}
\usepackage{multirow}

\tikzstyle{terminator} = [rectangle, draw, text centered, rounded corners, minimum height=2em]
\tikzstyle{process} = [rectangle, draw, text centered, minimum height=2em]
\tikzstyle{decision} = [diamond, draw, text centered, minimum height=2em]
\tikzstyle{data}=[trapezium, draw, text centered, trapezium left angle=60, trapezium right angle=120, minimum height=2em]
\tikzstyle{connector} = [draw, -latex']

\newcolumntype{L}[1]{>{\raggedright\let\newline\\\arraybackslash\hspace{0pt}}m{#1}}
\newcolumntype{C}[1]{>{\centering\let\newline\\\arraybackslash\hspace{0pt}}m{#1}}
\newcolumntype{R}[1]{>{\raggedleft\let\newline\\\arraybackslash\hspace{0pt}}m{#1}}

\title{
NLP-Powered Repository and Search Engine for Academic Papers: A Case Study on Cyber Risk Literature with CyLit
}

\date{}

\author[$\dag$]{Linfeng Zhang}
\author[$\star$]{Changyue Hu} 
\author[$\star$]{Zhiyu Quan} 

\affil[$\star$]{Program in Actuarial and Risk Management Sciences, University of Illinois Urbana-Champaign}

\affil[$\dag$]{Department of Mathematics, The Ohio State University}

\begin{document}
\maketitle
\begin{abstract}
As the body of academic literature continues to grow, researchers face increasing difficulties in effectively searching for relevant resources. Existing databases and search engines often fall short of providing a comprehensive and contextually relevant collection of academic literature. To address this issue, we propose a novel framework that leverages Natural Language Processing (NLP) techniques. This framework automates the retrieval, summarization, and clustering of academic literature within a specific research domain. To demonstrate the effectiveness of our approach, we introduce CyLit, an NLP-powered repository specifically designed for the cyber risk literature. CyLit empowers researchers by providing access to context-specific resources and enabling the tracking of trends in the dynamic and rapidly evolving field of cyber risk. Through the automatic processing of large volumes of data, our NLP-powered solution significantly enhances the efficiency and specificity of academic literature searches. We compare the literature categorization results of CyLit to those presented in survey papers or generated by ChatGPT, highlighting the distinctive insights this tool provides into cyber risk research literature. Using NLP techniques, we aim to revolutionize the way researchers discover, analyze, and utilize academic resources, ultimately fostering advancements in various domains of knowledge.

\vspace{1.0cm}


\end{abstract}

\section{Introduction}\label{sec:intro}

Literature databases and search engines play a crucial role in facilitating academic research and are indispensable resources for scholars across various disciplines. These resources offer valuable support to researchers, especially during literature reviews, by enabling them to explore pertinent studies within their respective fields, gain insights from previous research, identify seminal works, pinpoint research gaps, unearth potential avenues for future investigation, and contextualize their studies within the existing body of knowledge.

While these resources are crucial, it is important to acknowledge several significant caveats. First, current literature databases have limited coverage. Some of the most commonly used large literature databases, such as Web of Science\footnote{\url{https://www.webofscience.com/}} and Scopus\footnote{\url{https://www.scopus.com/}}, present a number of challenges. \citet{martin2018coverage} have noted a significant lack of consistency in literature coverage across various disciplines. These databases fail to provide access to a significant portion of highly cited literature in the fields of social sciences and humanities, ranging from 8.6\% to 28.2\%. The coverage limitations extend to specific forms of publications, including books and book chapters, which are not adequately represented in these databases. The analysis conducted by \citet{martin2016two} reveals that almost half of the 64,000 highly cited literature identified through Google Scholar is not listed in the Web of Science database, with approximately 18\% of the literature being books or book chapters. Furthermore, literature pertaining to specific domains might be scattered across various databases, necessitating a comprehensive search across multiple databases. However, manually conducting these searches is time-consuming and labor-intensive, presenting a notable challenge for researchers in need of efficient and all-encompassing access to pertinent academic resources. Second, search engines have a few noticeable drawbacks. Take Google Scholar\footnote{\url{https://scholar.google.com/}} as an example. Google Scholar is a leading academic search engine, and it excels in identifying research papers using a keyword-based search approach. It aims to imitate researchers when ranking literature, considering factors such as the full text of each document, the source of its publication, its authorship, and the frequency and recency of its citations within other scholarly publications. Google Scholar offers many advantages, including its capability to search for relevant books and articles in a single query, as well as its extensive coverage of books and conference proceedings. Nevertheless, it also comes with certain limitations. A notable limitation is the lack of domain-specific contextual awareness in keyword-based searches, leading to potential inaccuracy or irrelevance, particularly in interdisciplinary studies. For example, \textit{control}, as a polysemy, has various meanings depending on the context. Specifically, in the context of cyber risk, it refers to measures taken by an organization to enhance its cybersecurity. Google Scholar also does not provide users with the option to sort or search by academic discipline, and offers limited filtering options compared to conventional library databases. Furthermore, Google Scholar lacks transparency and clarity regarding search coverage, ranking methodology, and update frequency. Reverse engineering studies conducted by \citet{beel2009google} provide insights into this issue by showing that Google Scholar's ranking algorithm is heavily influenced by citation counts. Consequently, the search engine may have a preference towards commonly read literature, ignoring unconventional works or articles that present novel perspectives or viewpoints. Thus, \citet{beel2009google} emphasize the need for researchers to complement their search efforts with additional academic search engines or databases to ensure that their literature search is comprehensive and balanced.

The rapid acceleration of publication and innovation has triggered an exponential growth of academic literature across diverse research fields. Consequently, existing literature databases and search engines face challenges in coping with this surge, leading to an increasing demand among researchers for living literature reviews and academic search engines that cater to specific research domains. These living literature reviews offer numerous advantages, including enhanced search efficiency by minimizing irrelevant information, easy tracking of research trends to stay up-to-date with the latest developments, and the provision of a rich database for data mining, all of which foster new research insights. Moreover, they serve as shared knowledge platforms that encourage interdisciplinary collaboration and communication. To meet this increasing demand, a growing body of research is focusing on the development of tools tailored to specific research fields. For instance, \citet{wang2019visual} develop DNN Genealogy, an interactive visualization tool based on a systematic analysis of 140 publications that provides a visual summary of representative DNNs and their evolutionary relationships. Similarly, \citet{danilevsky2020survey} introduce XNLP, an interactive browser-based system that serves as a living literature review for cutting-edge research in Explainable AI (XAI) within the Natural Language Processing (NLP) domain. While these developments were initially accomplished manually by human reviewers, we propose employing NLP techniques to automate this process, enabling the processing of larger data volumes and further enhancing the efficiency of academic research. It should be noted that some web tools have been developed to automate the literature review process by implementing NLP techniques. 
In health and medical sciences, \citet{thomas2010eppi} introduce EPPI-Reviewer\footnote{\url{https://eppi.ioe.ac.uk/EPPIReviewer-Web/}}, a multi-user web application that streamlines the lifecycle of research synthesis reviews, allowing users to upload studies for screening, data extractions, and result analysis. \citet{bahor2021development} develop SyRF\footnote{\url{https://syrf.org.uk/}}, a fully integrated platform for conducting systematic reviews of preclinical studies, featuring automated bias item extraction for screening English articles. In environmental sciences, Colandr\footnote{\url{https://www.colandrapp.com/}} by \citet{cheng2018using} employs dual machine learning systems that not only rank articles by relevance but also categorize them by topic based on user input. Similarly, CADIMA\footnote{\url{https://www.cadima.info/}} by \citet{kohl_online_2018} facilitates systematic reviews, guiding users step-by-step through the review process, though it lacks built-in search and quantitative synthesis features. In software engineering, SESRA\footnote{\url{http://sesra.net/}} by \citet{sesra_2015} supports the complete systematic literature review process and is available in multiple languages. Additionally, in scientometrics, tools such as CiteSpace\footnote{\url{http://cluster.cis.drexel.edu/~cchen/citespace/}} \citep{chen2004searching, chen2006citespace, chen2010structure} and VosViewer\footnote{\url{https://www.vosviewer.com}} \citep{van2010software, eck2011text} are essential for conducting literature reviews, enabling the visualization and analysis of research trends and bibliometric networks within academic literature. For a comprehensive review of systematic literature review tools across different domains, see survey papers \citet{marshall2014tools}, \citet{o2015using}, \citet{feng2017text}, \citet{van2019software}, and \citet{harrison2020software}.

Although no NLP-based literature review tool has yet been developed in actuarial science, the application of NLP techniques has been explored, showcasing their significance in various contexts and their potential for being adopted for living literature reviews in this field. 
For instance, \citet{liao_text_2020} expand NLP use in insurance to customer service, employing text mining techniques such as topic modeling and sentiment analysis to analyze customer calls and improve operations in call centers. \citet{lee_actuarial_2020} incorporate text data into traditional insurance claim modeling by utilizing word similarity to extract risk features from claim descriptions, contributing to the improvement of insurance claims management and risk mitigation. Building on this, \citet{manski_extracting_2021} and \citet{manski_loss_2022} present a framework to predict the loss amount from textual descriptions of insurance claims using cosine similarities and word embedding. This framework utilizes automatic word selection instead of human-selected keywords, providing a more scalable and parsimonious model. \citet{zappa_text_2021} demonstrate the use of NLP in insurance by exploring how accident narratives from police reports can be used to classify risk profiles and fine-tune policy premiums. \citet{xu_bert-based_2022} adopt Bidirectional Encoder Representations from Transformer (BERT) to enhance the classification and severity prediction of truck warranty claims and demonstrate the superiority of BERT-based models in terms of accuracy and stability, highlighting the potential of NLP techniques such as BERT to improve predictive models in actuarial science.

To build on these advancements and leverage NLP techniques further, we propose a framework for retrieving, summarizing, and clustering relevant research papers within a specific field. The integration of NLP techniques enhances the system's capabilities by improving efficiency, extracting valuable insights from large volumes of unstructured text data, and refining the summarization of related literature.

Our main contributions are as follows:
\begin{enumerate}[label=(\arabic*)]
    \item We design and build a comprehensive framework that includes a living literature database and an academic search engine that caters to specific research domains. The proposed framework is equipped with state-of-the-art NLP techniques to enhance the effectiveness and efficiency of literature profiling and searching.
    \item We demonstrate the feasibility and practicality of this framework in the cyber risk domain and provide an unprecedented web tool\footnote{\url{https://cylit.math.illinois.edu/}} specifically designed for actuarial science researchers.
    \item Compared to the existing living literature review works, we have employed the most up-to-date NLP techniques, such as the newest BERT variants, for more reliable information extraction.
    \item We offer in-depth comparisons among the results generated by the proposed approach, human literature review, and large language models to highlight the advantages and limitations of all these methods. 
\end{enumerate}
To the best of our knowledge, this work is the first attempt at a living literature review in the actuarial science discipline, and it shows the potential of facilitating actuarial research in light of the rising volume of literature.

The remainder of this paper is structured as follows. In Section \ref{sec:method}, we expound on the methodology employed to create the proposed framework, emphasizing the utilization of NLP to accomplish our objectives. This section serves as a guide, offering insights into the creation and application of the framework. Section \ref{sec:cylit} offers a detailed exploration of the implementation of our proposed framework within the specific context of cyber risk. Here, we delve into practical aspects, showcasing how the framework operates in real-world scenarios with the accessible website. In Section \ref{sec:survey}, we undertake a comparative analysis, juxtaposing our proposed framework with the conventional survey papers typically conducted by human researchers. This comparative examination aims to highlight the distinctive features and advantages of our approach. Section \ref{sec:llm} offers an examination of ChatGPT's performance in paper categorization and literature review as opposed to the workflow consisting of manual review aided by the proposed literature search framework. The concluding Section \ref{sec:conclusion} summarizes the key findings and insights gleaned from our study. Additionally, it serves as a springboard for discussions of potential future directions in research and development within the scope of our proposed framework.

\section{Methodology}\label{sec:method}

\subsection{Intuition}\label{sec:int}

In the realm of comprehending and summarizing literature, human intuition typically involves a sequence of steps, starting with an initial assessment of the title and keywords, followed by perusing the abstract, delving deeply into the introduction and conclusion sections, and ultimately committing to a comprehensive reading of the entire literature. Inspired by this, we propose the utilization of NLP techniques to
imitate these steps taken by human readers
and automate the process of summarizing and categorizing literature. This method aims to enhance search efficiency, assist readers in tracking research trends, and offer novel insights for future research.  

In the interim, it is important to acknowledge potential challenges that NLP techniques may encounter. A well-crafted title should effectively convey the research topic, purpose, and scope while employing appropriate terminology and accurately reflecting the conducted work. However, titles may contain abbreviations, questions, or words intended to evoke interest, which can confuse NLP techniques when attempting to extract relevant information. For instance, the highly cited NLP paper titled ``Attention is all you need'' \citep{vaswani2017attention} may have a captivating title, but from a text-mining perspective, only the term ``attention" might be useful for summarization purposes. Keywords are essential for capturing the essence of literature and for identifying its research focus. Authors can enhance the searchability of their work by incorporating relevant keywords. Typically, keywords consist of 2-4 word phrases, with each paper summarized by 3-5 keywords. It is common for related studies to share similar keywords. However, not all papers provide explicit keywords. For instance, the aforementioned paper, ``Attention is all you need'', does not include keywords. In such cases, a careful examination of the abstract becomes necessary to extract relevant information. Abstracts, usually limited to around 350 words, present a summary of the paper's main points, including the research problem, the basic design of the study, key findings resulting from the analysis, and concise conclusions. The writing styles of abstracts vary across disciplines, posing challenges when summarizing and comparing abstracts to identify related papers. Nevertheless, abstracts remain valuable, as they allow authors to elaborate on key aspects of their work, often yielding more information than keywords alone. The Introduction section serves the purpose of guiding readers from a broad subject area to a specific research field. It establishes context by summarizing existing knowledge, providing background information, stating the purpose of the study, and briefly outlining the authors' rationale, methodology, potential outcomes, and the paper's overall structure. On the other hand, the conclusion section summarizes the paper and synthesizes its key points. Both the introduction and the conclusion are more extensive than the abstract and contain more detailed information, which can be utilized if the abstract is insufficient.

\begin{figure}[h!]
  \centering
  \includegraphics[width= 1.0\linewidth]{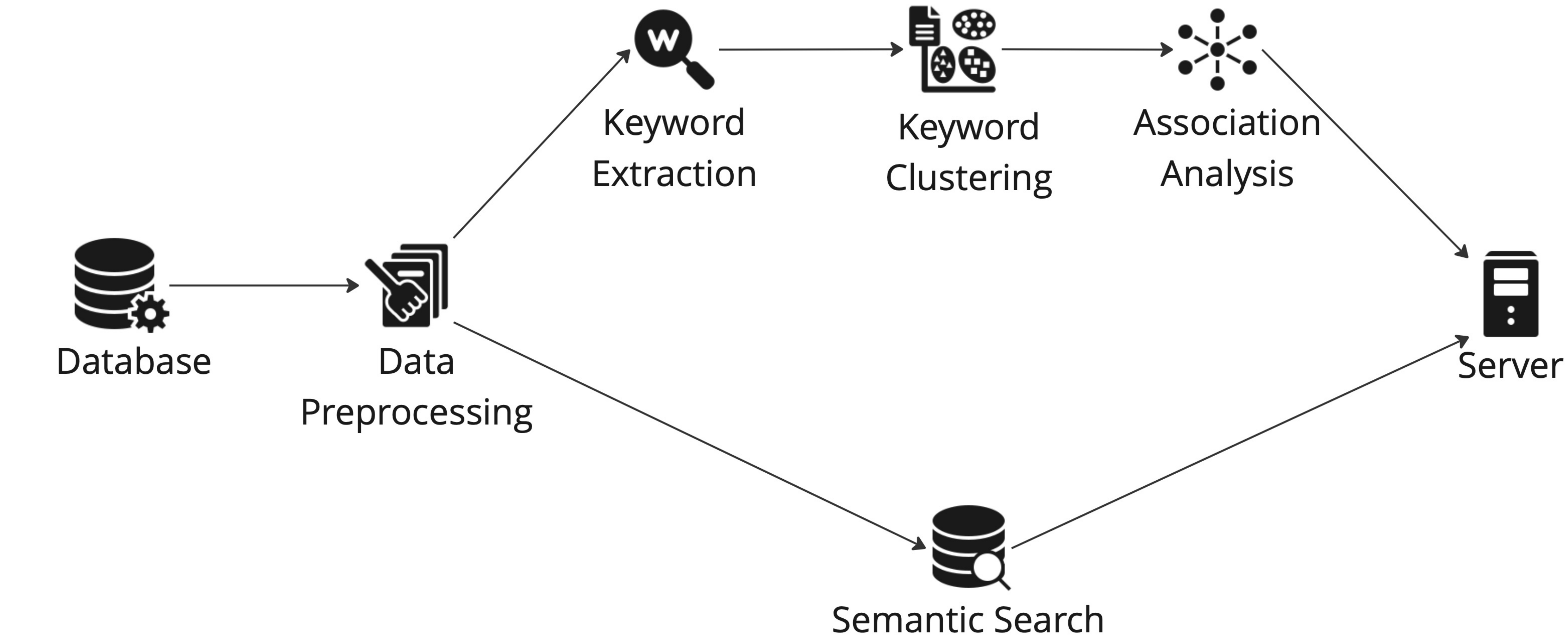}
  \caption{NLP-powered literature system}
  \label{fig:nlp}
\end{figure}

Considering the aforementioned key elements and conducting several rounds of experiments, our primary approach involves leveraging keyword information to summarize papers into concise phrases. By categorizing papers into clusters based on these keywords, we gain valuable insight into emerging research trends and identify potential interdisciplinary activities. In cases where keywords are not provided, we utilize additional textual information from the title, abstract, introduction, and conclusion to generate appropriate keywords for the paper. Furthermore, our search engine combines the aforementioned key elements to obtain summarized information that best matches the query information. Figure \ref{fig:nlp} presents the chain of steps in our NLP-powered literature system.
\subsection{Word Embedding}\label{sec:nlp}

NLP serves as a valuable tool for transforming raw unstructured text information into structured data suitable for analysis. In our paper, we present a concise overview of the NLP techniques we have examined. An essential aspect of NLP involves the conversion of text into a numerical representation that computers can comprehend.

The term \emph{word embedding} refers to representing words for text analysis in the form of a real-valued vector that encodes the meaning of words, resulting in words with similar meanings being closer in the vector space. Word embeddings can be obtained using a set of NLP techniques where words or phrases from the vocabulary are mapped to vectors of real numbers. Our study mainly focuses on exploiting the mathematical properties of word embeddings and how they interact in an $n$-dimensional vector space. In this study, we investigate several methods to generate word embeddings: Term Frequency Inverse Document Frequency (TF-IDF), Word2vec \citep{mikolov2013efficient, mikolov2013distributed}, Bidirectional Encoder Representations from Transformers (BERT) \citep{devlin2018bert}, and derivations of BERT, Sentence-BERT \citep{reimers2019sentence} as well as KeyBERT \citep{grootendorst2020keybert}.

Commencing with the conventional approach, TF-IDF is a numerical statistic that measures the relevancy of a word in relation to a collection of documents. Term frequency represents the number of times a word appears in a document, while the inverse document frequency indicates the frequency of a word in the entire collection of documents. The TF-IDF value increases proportionally with the number of times a word appears in the document, but is offset by the number of documents in the corpus that contain the word, thereby adjusting for the disruption caused by some words appearing more frequently in general. TF-IDF has become one of the most popular term-weighting schemes and can be used to generate basic summary statistics to identify significant keywords. However, TF-IDF alone may not be sufficient for our task, primarily due to its limitation as the number of documents grows, the size of the embeddings grows exponentially along with it, resulting in a loss of information and an escalation of noise within the data. Consequently, a bag-of-words (BoW) approach may not be the most suitable option for a large corpus. 

To circumvent this problem, we investigate more sophisticated word embedding techniques. One such method is Word2vec, which comprises a family of models that employ shallow neural networks to generate word embeddings and capture word associations from an extensive corpus of text. In Word2vec, linguistic contexts are reconstructed using either continuous bag-of-words (CBOW) or continuous skip-gram architectures. In the CBOW architecture, the model predicts the current (middle) word by using the surrounding context words within a specified window. The context consists of a few words before and after the current (middle) word. Conversely, in the continuous skip-gram architecture, the model uses the current word to predict the surrounding window of context words. In other words, it predicts words within a certain range before and after the current word in the same sentence. Through the large corpus of linguistic context reconstruction (model training) process, Word2vec represents each distinct word with vectors, typically containing several hundred dimensions, that capture words' semantic and syntactic qualities. Ideally, these word vectors are positioned in the vector space such that words that share common contexts in the corpus, \textit{i.e.}, semantically and syntactically similar, are located close to one another. Conversely, more dissimilar words are placed farther apart. Hence, the degree of semantic and syntactic similarity between words represented by vectors can be measured by a simple mathematical function, \textit{e.g.}, cosine similarity. However, Word2vec may be suboptimal since it relies on local information. In other words, since the semantic and syntactic representation of a word relies only on its neighbors, it cannot comprehend words under the big picture of the document. This phenomenon is not suitable for academic papers or scientific articles. 
In addition, since Word2vec assigns one-to-one relationships to words and vectors, it does not solve the problem of polysemous words. For example, the aforementioned word \textit{control} has a specific meaning in cybersecurity, and thus its vector representation ought to be close to that of the word \textit{security} if the context is correctly comprehended. However, without this semantic context, the distance between these two words may be incorrectly represented.
Moreover, using pre-trained models that are not specifically designed for the target domain can lead to inaccurate results. For instance, a pre-trained Word2vec model using the Google News dataset may not generalize well to the domain of cyber risk. Furthermore, Word2vec struggles with out-of-vocabulary words, as it generates word embeddings based on its training data and assigns random vector representations to out-of-vocabulary words.

BERT belongs to the family of transformer-based models and is designed to understand the contextual meaning of words in a sentence. BERT is a pre-trained deep (learning) bidirectional representation using a large amount of unlabeled text data from diverse sources, such as books, articles, and web pages. Its bidirectional training allows BERT to learn the contextual representation of words by considering the entire sentence, surpassing the limitations of traditional directional approaches that only consider neighboring words. This bidirectional approach allows BERT to acquire a deeper understanding of word relationships and contextual nuances. BERT is pre-trained on two tasks: language modeling and next-sentence prediction. In language modeling, BERT has been trained to predict randomly masked words from the surrounding context, with approximately 15\% of the words being masked. In next-sentence prediction, BERT is trained to determine whether a given second sentence is likely to follow a given first sentence, since language modeling alone does not inherently capture the relationship between two sentences. As a result of the training process, BERT learns contextual embeddings for words. The pre-trained BERT model can be further fine-tuned by adding just one additional output layer to create state-of-the-art models for a wide range of downstream tasks, such as text classification, named entity recognition, question answering, and more, where BERT has demonstrated exceptional performance.
 
Word2vec generates a single-word embedding representation for each word in the corpus. For example, the word ``attention" has the exact Word2vec vector representation in both sentences ``Attention is all you need'' and ``Please pay attention''. In contrast, BERT offers contextualized embeddings that vary based on the sentence. Since the nature of a sequential input, BERT considers the context for each occurrence of the given word and allows the word embeddings to store contextual information. BERT also effectively addresses the out-of-vocabulary. BERT learns at the subword level, \textit{i.e.}, instead of learning and processing entire words, BERT breaks down words using \textit{WordPiece tokenization} into smaller units called subwords or tokens. This gives members of the BERT family a smaller vocabulary than the initial training data. Because of this, BERT can generate embeddings for out-of-vocabulary words, giving it an expansive vocabulary. Therefore, BERT is better suited for our purpose compared to Word2vec. 

As our literature collection continues to grow, the complexity of time and calculations increases exponentially. On our website, we require extensive semantic text searches and similarity clustering, which poses challenges due to the computational overhead associated with traditional BERT. In search of an alternative model, we have discovered Sentence-BERT, a modified version of pre-trained BERT specifically designed to generate semantically meaningful sentence embeddings. These embeddings can be compared using cosine similarity, offering a more efficient solution for identifying the most similar sentences within a large collection. Sentence-BERT significantly reduces computation time, making it well-suited for our needs in handling the size of our current literature corpus and accommodating anticipated future growth. Among the BERT family of models, Sentence-BERT has emerged as the most suitable approach for our requirements.

\subsection{Keyword Extraction and Clustering}\label{sec:keyword_extraction_and_clustering}
This section delves into a comprehensive exploration of utilizing keyword information to establish clusters, enabling a deeper understanding of current research trends and the interconnections among literature in specific fields of study. It comprises three key steps: keyword extraction, keyword clustering, and association analysis between keyword clusters. We consider two scenarios: one where authors provide keywords and another where they do not. When authors provide keywords, we preprocess and consolidate them to form a domain-specific keyword library. In cases where keywords are not provided, we preprocess the abstracts and apply keyword extraction techniques to identify relevant keywords from the cleaned abstracts. To ensure accuracy and minimize errors in keyword extraction, we cross-reference the extracted keywords with those provided by the authors. The intersection between the extracted keywords and the existing comprehensive keyword library serves as the final keyword selection. Once each paper is associated with a set of keywords, we perform keyword clustering using the keyword library to uncover topics within the research area. Additionally, we conduct association analysis among the identified topics to reveal cross-topic research activities. Each step and the methodologies used will be detailed in the subsequent parts of this section.

\subsubsection{Preprocessing}\label{sec:preprocessing}

Raw text data often have many undesirable characteristics that make it difficult for NLP models to process. For example, ``cybersecurity'' might be written as ``Cybersecurity'', ``cyber-security'', or ``cyber security'' depending on the sources of the text data. Although all the variants have the same meaning, their different formats are essentially noises, making it difficult for the machine to interpret. Therefore, preprocessing procedures that fix these inconsistencies and reduce noises are crucial for effective text analysis.
In our approach, we perform preprocessing steps on the author-provided keywords, if available; otherwise, the abstract is preprocessed to create a standardized set of keywords and a clean, structured dataset of abstracts. These preprocessed datasets can then be efficiently utilized for further analysis and modeling.

To standardize the provided keywords, we implement a series of preprocessing steps. Initially, we remove any punctuation or special characters that may introduce discrepancies or hinder keyword matching. This step helps eliminate potential noise in the data, ensuring a cleaner set of keywords suitable for analysis. Next, we convert all keywords to lowercase to maintain consistency throughout the dataset. Furthermore, we address duplicate entries by removing spaces between certain key phrases and expanding specific abbreviations to their full forms. This reduces the dimensionality of the dataset and ensures that semantically similar keywords are not treated as distinct entities. To maintain consistency, we apply the same preprocessing procedures to the abstracts. This allows us to identify and group words that possess similar meanings but may be written differently. Through this preprocessing process, we standardize both the keywords and abstract datasets, facilitating more effective analysis and comparison across papers.

\subsubsection{Keyword Extraction}\label{sec:keyword_extraction}

\begin{figure}[h!]
  \centering
  \includegraphics[width= 1.0\linewidth]{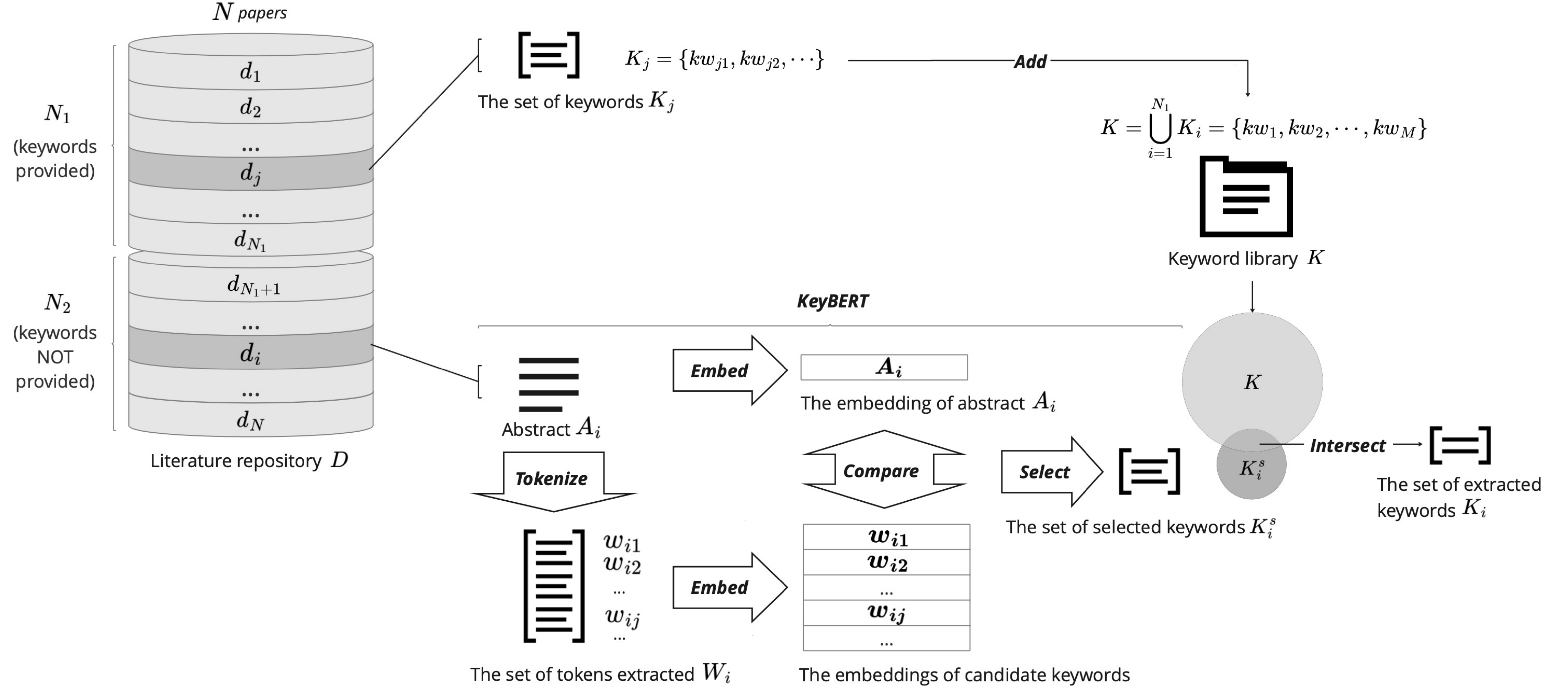}
  \caption{The workflow of keyword extraction algorithm}
  \label{fig:extraction_pipe}
\end{figure}

We present the workflow of keyword extraction in Figure \ref{fig:extraction_pipe}. Consider that we have acquired a literature repository, denoted by $D$, which focuses on a particular research domain. The repository contains $N$ papers, represented by $D = \{d_1, d_2, \cdots, d_N\}$, where $i=1, \cdots, N$. Among these papers, the first $N_1$ of them include author-provided keywords, while the remaining $N_2$ papers require keyword extraction. The total number of papers is the sum of $N_1$ and $N_2$ (\textit{i.e.}, $N = N_1 + N_2$). Each paper, denoted by $d_i$, is accompanied by an abstract denoted by $A_i$ and a set of keywords, if provided by the authors, denoted by
$K_{i} = \{kw_{i1}, kw_{i2}, \cdots\}$. Note that $K_i = \emptyset$ when $i=N_1 + 1, \cdots, N$.

Following the preprocessing procedures detailed in Section \ref{sec:preprocessing}, we collect the keywords provided by the author, $K_{i}$, and consolidate them to create an initial keyword library, denoted by $K =\bigcup_{i=1}^{N_1} K_{i}=\{kw_{1}, kw_{2}, \cdots, kw_{M}\}$, where $M$ denotes the total number of keywords in the library. We consider $K$ as a comprehensive and representative keyword library when $N$ is substantially large and $N_1 \gg N_2$. We perform keyword extraction in the abstract for the $N_2$ papers. Although we could expand the keyword extraction input text by including the introduction, conclusion, or even entire papers, our paper concentrates solely on the abstract. This decision is based on two factors: first, including more content than the abstract only offers limited improvements; and second, there is a consideration for computational efficiency along with copyright, licensing, and access to full-text content for text mining. There are two schemes for keyword extraction. The first scheme involves utilizing the statistical properties of the bag-of-words (BoW) approach, exemplified by methods like Rake \citep{rose_rake_2010} and YAKE \citep{campos_yake_2020}. The second scheme leverages pre-trained embeddings, such as KeyBERT. 

KeyBERT, introduced by \citet{grootendorst2020keybert}, is a state-of-the-art keyword extraction technique that leverages BERT embeddings to extract the most semantically relevant keywords (or key phrases) from documents. The KeyBERT algorithm involves several critical steps: \emph{tokenization}, \emph{embedding}, and \emph{selection}.

In the initial stage, given an abstract $A_i$ for $i \in \{N_1 +1, \cdots, N\}$, and a pre-specified range of \texttt{n-gram} (contiguous sequences of $n$ words), KeyBERT uses a vectorizer, such as \texttt{CountVectorizer} from the Python \texttt{scikit-learn} library, to tokenize $A_i$ into a set of \texttt{n-gram} candidate keywords or key phrases, 
\begin{equation*}
    W_i = \{w_{i1}, w_{i2}, \cdots, w_{ij}, \cdots \}, \text{ for } ij = i1, i2, \cdots
\end{equation*}
where $w_{ij}$ represents the $j$-th token (word or phrase) extracted from $A_i$. 

Let $E(\cdot)$ denote the embedding function. Subsequently, KeyBERT computes the embeddings for both $A_i$ and the candidate keywords in $W_i$ using the pre-trained Sentence-BERT model,  
\begin{align*}
E(A_i) &= \boldsymbol{A_i}, \text{ for } i = N_1 + 1, \dots, N,\\
E(w_{ij}) &= \boldsymbol{w_{ij}}, \text{ for } ij = i1, i2, \cdots
\end{align*}
where $\boldsymbol{A_i}$ and $\boldsymbol{w_{ij}}$ are vectors representing the embeddings for $A_i$ and the $j$-th candidate keyword, respectively. KeyBERT also allows the use of seed keywords to guide keyword extraction by steering similarities toward the seed keywords. In cases where seed keywords are provided, the KeyBERT algorithm will modify the document embedding by computing the weighted average of the previous document embedding and the seed keyword embeddings.

In the most important selection step, KeyBERT chooses the $m$ most representative keywords from the candidate set $W_i$ based on one of the following methods: Cosine Similarity, Maximal Marginal Relevance (MMR), and Max Sum Distance.
\begin{itemize}
\item \textbf{Cosine Similarity}: This method computes the cosine similarity between the embedding of the candidate keyword, $\boldsymbol{w_{ij}}$, and the embedding of the abstract, $\boldsymbol{A_i}$,
\begin{equation*}
    \text{sim}(\boldsymbol{w_{ij}}, \boldsymbol{A_i}) = \frac{\boldsymbol{w_{ij}} \cdot \boldsymbol{A_i}}{\|\boldsymbol{w_{ij}}\| \|\boldsymbol{A_i}\|}.
\end{equation*}
Then, keyBERT selects $m$ keywords that maximize their cosine similarities with document embedding.
\item \textbf{Maximal Marginal Relevance}: 
MMR, introduced by \citet{carbonell1998use}, is used in KeyBERT to balance the diversity and relevance of selected keywords by maximizing both the dissimilarity among these keywords and their similarities to the document. MMR gives a set of selected keywords based on the following criterion, 
\begin{equation*}
    \arg \max_{w_{ij} \in W_i\setminus K^s_i} \left[ \left(1 - \alpha \right)\text{sim}\left( \boldsymbol{w_{ij}}, \boldsymbol{A_i} \right) - \alpha \max_{w^s_{ik} \in K^s_i} \text{sim}(\boldsymbol{w_{ij}}, \boldsymbol{w^s_{ik}}) \right],
\end{equation*}
where $K^s_i$ is the set of selected keywords, $\boldsymbol{w^s_{ik}}$ represents the embedding of the $k$-th selected keyword $w^s_{ik}$, and $\alpha$ is the diversity parameter. Note that $\alpha$ can be fine-tuned using papers with author-provided keywords as training data. Starting from an empty set of selected keywords, of which the cardinality $|K^s_i| = 0$, the maximization algorithm iterates to pick unselected candidate keywords until $|K^s_i| = m$.
\item \textbf{Max Sum Distance}: Alternatively, the Max Sum Distance method aims to maximize the sum of pairwise distances among the selected keywords. It begins by computing the cosine similarities between the embedding of each candidate keyword and the document embedding, $\text{sim}(\boldsymbol{w_{ij}}, \boldsymbol{A_i})$, and then takes $2m$ candidate keywords that are most relevant to the paper based on the highest cosine similarity values,
\begin{equation*}
    W_i^c = \{ w_{i1}^c, w_{i2}^c, \cdots, w_{ij}^c, \cdots \}, \quad \left|W_i^c \right| = 2m.
\end{equation*}
Then it iterates through all the possible combinations of $m$ candidates from the set of selected candidates $W_i^c$ to find the combination that has the lowest sum of pairwise similarities (or the highest sum of pairwise distances),
\begin{equation*}
    \arg\min_{K^s_i\subseteq W_i^c, |K^s_i| = m} \sum_{(w_{ij}^c, w_{ik}^c) \in W_i^c \times W_i^c, j \neq k} \text{sim}(\boldsymbol{w_{ij}^c},\boldsymbol{w_{ik}^c}),
\end{equation*}
where $\boldsymbol{w_{ij}^c}$ and $\boldsymbol{w_{ik}^c}$ represent the embeddings of the candidate keywords $w_{ij}^s$ and $w_{ik}^s$ respectively.
\end{itemize}
Lastly, the keyBERT algorithm outputs the $m$ selected keywords and keyphrases along with their respective similarity scores with document embedding,
\begin{equation*}
    K^s_i = \{w_{i1}^s, w_{i2}^s, \cdots, w_{ij}^s, \cdots w_{im}^s\}.
\end{equation*}
To determine the final selection of keywords, we update $K_i$ by identifying the intersection between the extracted keywords from KeyBERT and the library of preprocessed keywords provided by the authors,
\begin{equation*}
K_i = K^s_i \cap K, \text{ for } i = N_1 + 1, \dots, N.\\
\end{equation*}
Using this approach, we extract precise and diverse keywords that effectively capture the semantic essence of papers.

\subsubsection{Keyword Clustering}\label{sec:keyword_cluster}

\begin{figure}[h!]
  \centering
  \includegraphics[width= 1.0\linewidth]{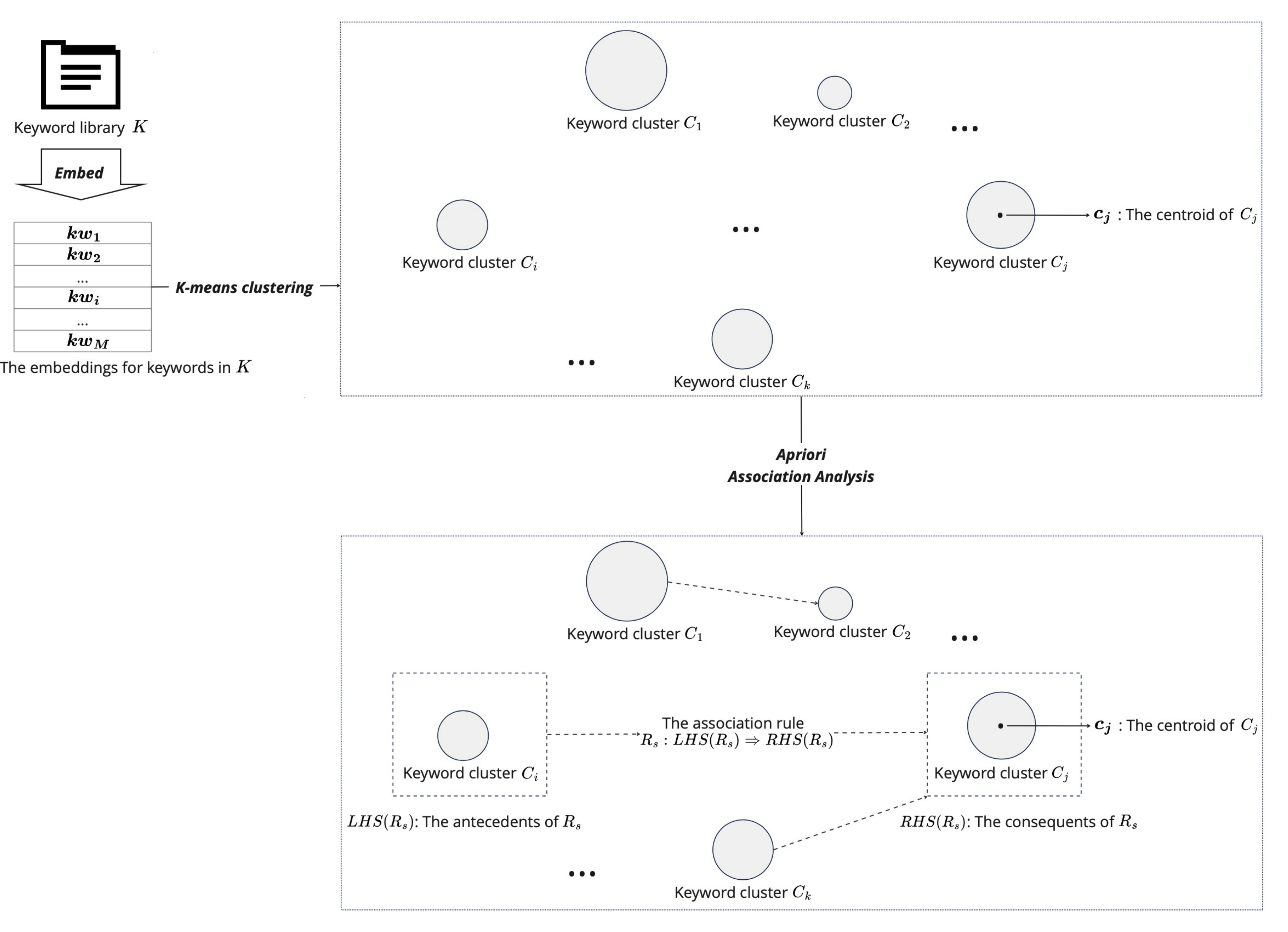}
  \caption{The workflow of keyword clustering and association analysis between clusters}
  \label{fig:clustering_pipe}
\end{figure}

Keyword clustering groups similar keywords together, which helps summarize the keyword library and identify the main topics discussed within a specific research domain. We illustrate the pipeline of keyword clustering and the subsequent association analysis in Figure \ref{fig:clustering_pipe}. The K-means algorithm, first introduced by \citet{macqueen1967classification}, is one of the most well-known clustering algorithms. It has been widely applied to textual data for identifying potential topics. Given a predefined number of clusters $k$, K-means clustering aims to partition the keywords into $k$ distinct clusters so that the distances within each cluster are minimized. Given a set of keywords $K = \{kw_1, kw_2, \cdots, kw_M\}$, we first compute the embedding of each keyword using the same pre-trained Sentence-BERT model used in Section \ref{sec:keyword_extraction},
\begin{equation*}
    \{\boldsymbol{kw_1}, \boldsymbol{kw_2}, \cdots, \boldsymbol{kw_i}, \cdots, \boldsymbol{kw_M}\},
\end{equation*}
where $\boldsymbol{kw_i}$ represents the embedding of the $i$-th keyword in the keyword library. The K-means algorithm consists of several steps: \emph{centroid initialization}, \emph{cluster assignment}, \emph{centroid update}, and \emph{iterative optimization}.

\emph{Centroid initialization}. The K-means algorithm begins by randomly selecting $k$ initial centroids, 
\begin{equation*}
    \{\boldsymbol{c_1}, \boldsymbol{c_2}, \cdots, \boldsymbol{c_j}, \cdots, \boldsymbol{c_k}\}
\end{equation*}
where each centroid $\boldsymbol{c_j}$ represents the mean vector of the $j$-th keyword cluster. 

\emph{Cluster assignment}. Then K-means assigns each keyword $kw_i$ to the nearest centroid $\boldsymbol{c_j}$ by calculating the Euclidean distance between the keyword and each centroid,
\begin{equation*}
C_j = \{kw_i: ||\boldsymbol{kw_i} - \boldsymbol{c_j}||^2 \leq ||\boldsymbol{kw_i} - \boldsymbol{c_l}||^2, \forall l, 1 \leq l \leq k\},
\end{equation*}
where $C_j$ denotes the $j$-th keyword cluster and $||\cdot||$ represents the Euclidean distance.

\emph{Centroid update}. Subsequent to the cluster assignment, the centroids $\boldsymbol{c_j}$ are recalculated by averaging the embeddings of all keywords in cluster $C_j$:
\begin{equation*}
\boldsymbol{c_j} = \frac{1}{|C_j|} \sum_{{kw_i}\in C_j} \boldsymbol{kw_i},
\end{equation*}
where $|C_j|$ is the number of keywords in the cluster $j$.

\emph{Iterative optimization}. The \emph{cluster assignment} and \emph{centroid update} steps are iterated until a predefined stopping criterion is met. The predefined stopping criterion can be the maximum number of iterations or the convergence of objective function, such as the within-cluster distances,
\begin{equation*}
\sum_{j=1}^{k} \sum_{kw_i \in C_j} ||\boldsymbol{kw_i} - \boldsymbol{c_j}||^2.
\end{equation*}
By implementing the K-means algorithm, we are able to generate keyword clusters along with their corresponding centroids. This allows us to identify the topics associated with these keywords effectively. Visualization techniques such as word clouds can be used to create a visually appealing and easily understandable representation of each topic. Such graphical representations facilitate the understanding of the derived topics, making the presentation of the findings more accessible.

\subsubsection{Association Analysis}\label{sec:association}

From keyword clustering analysis, we categorize semantically similar keywords and subsequently identify papers that cover similar topics. It is noteworthy that a paper may include keywords from multiple topics simultaneously. For instance, consider a paper with a set of keywords $K_i = \{kw_{i1}, kw_{i2}, kw_{i3}\}$, where $kw_{i1}$ and $kw_{i2}$ belong to the keyword cluster $C_1$ and $kw_{i3}$ belongs to keyword cluster $C_2$. Then the paper is categorized under a set of keyword clusters $\{C_1, C_2\}$. There may be a notable co-occurrence of the clusters $C_1$ and $C_2$ in the keyword sets across our collection of papers. Such frequent co-occurrence of specific topics in papers may suggest a potential association between them. To gain insight into cross-topic research activities, we investigate the association patterns among keyword clusters $\{C_1, C_2, \cdots, C_k\}$. Association rule mining is widely used to reveal the underlying connections between different items. Among various techniques for association rule mining, the Apriori algorithm, introduced by \citet{agrawal1994fast}, is one of the most popular techniques. In the actuarial science literature, \citet{risks6030069} use the Apriori algorithm to discover empirical evidence of a potential association between the policyholder-switching following a claim and the consequent change in premium. The algorithm measures the strength of the relationships between items based on three key metrics, namely, \emph{support}, \emph{confidence}, and \emph{lift}. Given a set of association rules $\{R_1, R_2, \cdots, R_s, \cdots \}$, each rule $R_s$ specifies an association between two or more keyword clusters,
\begin{equation*}
    R_s: LHS(R_s) \Rightarrow RHS(R_s), 
\end{equation*}
where $LHS(R_s) = \left\{ C_{\mathcal{U}} \mid \mathcal{U} \subseteq  \{1, 2, \cdots, k \} \right\}$ are the antecedents of $R_s$, and $RHS(R_s) = \{ C_\mathcal{V} \mid \mathcal{V} \subseteq  \{1, 2, \cdots, k \} \}$ are the consequents of $R_s$. 
In other words, if we observe keyword clusters in $LHS(R_s)$ from the paper's keywords, it implies the presence of a keyword cluster in $RHS(R_s)$ as per the association rule $R_s$. It is important to note that $LHS(R_s)$ and $RHS(R_s)$ are mutually exclusive sets of keyword clusters.
The \emph{support} of a set of keyword clusters $LHS(R_s)$ is defined as the proportion of papers that contain keywords from all the clusters in $LHS(R_s)$,  
\begin{equation*}
\emph{support}(LHS(R_s)) = \frac{N_{\mathcal{U}}}{N},
\end{equation*}
where $N_{\mathcal{U}}$ is the number of papers containing keywords from all the clusters in $LHS(R_s)$.
The \emph{confidence} of a rule $R_s$ measures the proportion of papers containing keywords from all the clusters in $R_s$ among those papers that contain the keywords from all the clusters in the antecedent of $R_s$, 
\begin{equation*}
\emph{confidence}(R_s) = \frac{\emph{support}(R_s)}{\emph{support}(LHS(R_s))},
\end{equation*}
The \emph{lift} of rule $R_s$ measures the degree of dependence between the antecedents and the consequents of an association rule, considering the frequency of co-occurrence of the keyword clusters. The lift of $R_s$ is defined as
\begin{equation*}
\emph{lift}(R_s) = \frac{\emph{confidence}(R_s)}{\emph{support}(RHS(R_s))} = \frac{\emph{support}(R_s)}{\emph{support}(LHS(R_s))\cdot\emph{support}(RHS(R_s))}
\end{equation*}
A lift value larger than $1$ indicates a positive association between the antecedents and the consequents of the rule, suggesting that the occurrence of one improves the likelihood of the occurrence of the other. Conversely, a lift value smaller than $1$ indicates a negative association, the occurrence of one negatively impacts the occurrence of the other. When the lift value is exactly $1$, it implies independence between the antecedents and the consequents, indicating that there is no noticeable association. The association rules obtained from the Apriori analysis are filtered using predefined criteria based on the aforementioned metrics to the relevance and significance of the extracted associations.

\subsection{Semantic Search}\label{sec:faiss}
Semantic search refers to the process by which search engines endeavor to comprehend the contextual meaning of a user's search query, aiming to return results that align with the searcher's intent. In contrast to lexicographical search methods that seek exact matches, semantic search focuses on grasping the meaning and context of the query. For instance, in our semantic search system, if the user searches ``how much does a cyber attack cost?", the system is able to return a paper titled ``Insuring against cyber-attacks," which discusses cyber insurance, even though the word ``insurance" is not present in the query. On the other hand, a basic lexicographic search might not return any useful information since the search function cannot comprehend the meaning behind the question. This approach offers the advantage of identifying pertinent literature that may not strictly match the query terms but is semantically connected.

\begin{figure}[h!]
  \centering
  \includegraphics[width= 1.0\linewidth]{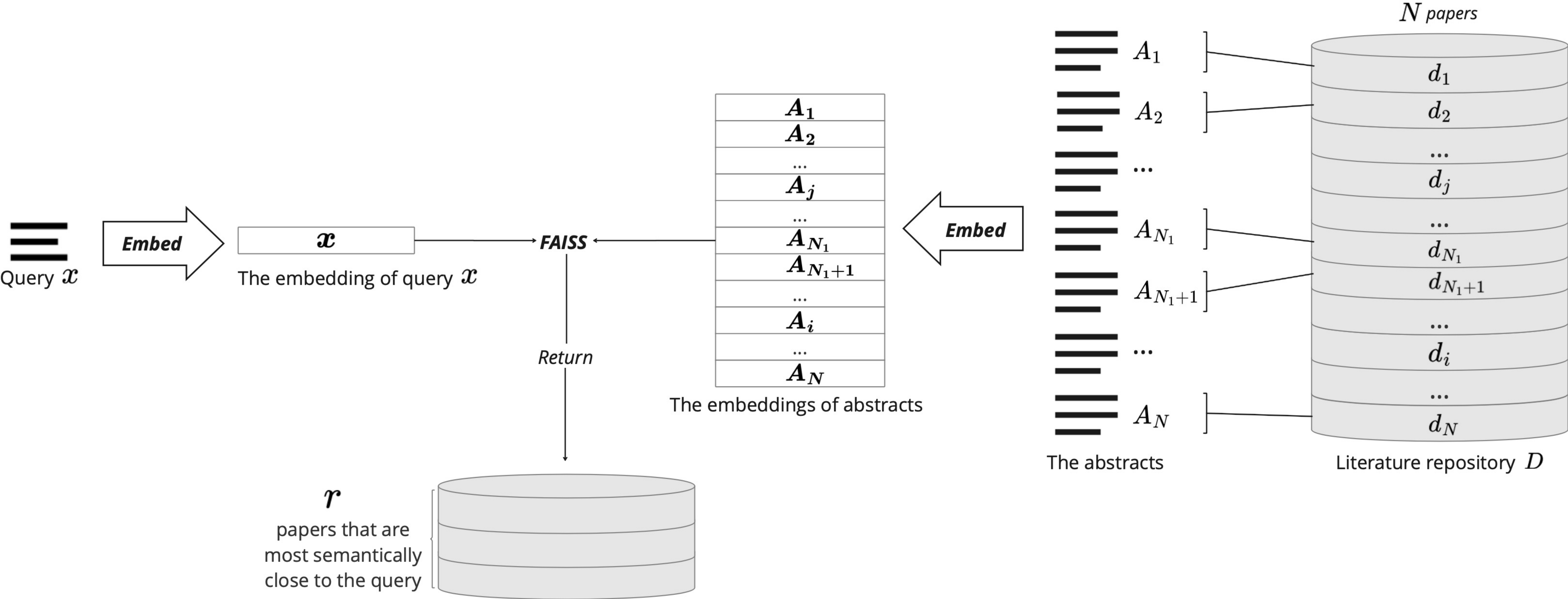}
  \caption{The workflow of semantic search}
  \label{fig:search_pipe}
\end{figure}

We illustrate the workflow of semantic search in Figure \ref{fig:search_pipe}. As discussed in the previous section, we convert keywords and abstracts into vectors using Sentence-BERT models, which allows us to use similarity metrics such as cosine similarity to identify abstracts of papers that demonstrate semantic similarity to the query phrase. Nevertheless, this approach presents some challenges, especially as the size of the database increases. With an increasing number of papers, the system must undertake a growing number of pairwise similarity comparisons across the entire database. In addition, the volume of comparison tasks increases with the number of concurrent semantic search users.

Performing similarity searches on a large scale may pose two primary challenges. First, some conventional techniques require loading the entire vector set into the system memory, which may not be sufficient for handling large data sets. Second, ensuring the efficient execution of search processes and the timely delivery of search results becomes an exceedingly challenging task. To tackle these concerns, we utilize the Facebook AI Similarity Search (FAISS) indexing method for text vectors, which offers an effective solution for managing large-scale semantic searches.

FAISS, introduced by \citet{johnson2019billion}, provides an efficient solution for comparison and similarity searches of high-dimensional vectors. The core of the algorithm consists of several stages: \textit{vector quantization}, \textit{index building}, and \textit{efficient similarity search}. We have a set of $N$ Sentence-BERT embedding vectors,
\begin{equation*}
    \{\boldsymbol{A_1}, \boldsymbol{A_2}, \cdots, \boldsymbol{A_i}, \cdots, \boldsymbol{A_N}\} \in \mathbf{R}^p
\end{equation*}
where $\boldsymbol{A_i}$ represents the embedding of the $i$-th abstract and $p$ is the number of dimensions for embedding vectors. Given a query $x$ and its embedding $\boldsymbol{x} \in \mathbf{R}^p$, the goal is to identify the $r$ papers that are semantically closest to the query $x$. 

In the \emph{vector quantization} stage, high-dimensional vectors are mapped to a lower-dimensional space using quantizers. A quantizer function $q$ maps a $p$-dimensional vector $\boldsymbol{y} \in \mathbf{R}^p$ to its nearest centroid in a codebook $\mathcal{C} = \{\boldsymbol{c_i}: i \in \mathcal{I}\}$, with $\mathcal{I} = \{1, 2, \cdots, |\mathcal{C}|\}$. The codebook size is $|\mathcal{C}|$, and quantizers are generally trained via K-means clustering. Product quantization (PQ), introduced by \citet{jegou_product_2010}, further splits the vector $\boldsymbol{y}$ into $b$ subvectors $\boldsymbol{y}=\left[\boldsymbol{y^1}, \cdots, \boldsymbol{y^b}\right]$ of dimension $p/b$. Each subvector is quantized separately to generate
$$ \left(q^1(\boldsymbol{y^1}), \cdots, q^j(\boldsymbol{y^j}), \cdots, q^b(\boldsymbol{y^b}) \right),$$ 
where $q^j$ is the quantizer for the $j$-th subvector of $\boldsymbol{y}$.

The \emph{index building} stage involves constructing an Inverted File with Asymmetric Distance Computation (IVFADC) index, facilitating efficient similarity search. FAISS manages memory usage efficiently by maintaining an index file on the hard disk. This index file is used to construct a significantly smaller data structure in computer memory, thereby addressing memory insufficiency issues. FAISS creates the index file by partitioning the database into multiple clusters based on quantized vector centroids.  
It employs a two-level quantization approach,
$$q_1(\boldsymbol{A_i}) + q_2(\boldsymbol{A_i} - q_1(\boldsymbol{A_i})).$$
The first level of quantization, denoted as $q_1$, is known as the coarse quantizer. It categorizes vectors into different clusters, effectively partitioning the dataset. Correspondingly, an inverted file, a data structure that groups the vectors $\boldsymbol{A_i}$ into $\left|\mathcal{C}_1\right|$ inverted lists 
with homogeneous $q_1 \left(\boldsymbol{A_i}\right)$, is maintained. The codebook size $\left|\mathcal{C}_1\right|$ is typically around $\sqrt{N}$. The second level of quantization, denoted as $q_2$, is referred to as the fine quantizer. It encodes the remaining information after the first-level coarse quantization, providing a detailed representation of the vectors within each cluster. The fine quantizer $q_2$ is a product quantizer with $b$ subquantizers,
$$
\left(q_2^1\left(\boldsymbol{A_i}-q_1\left(\boldsymbol{A_i}\right)\right), \cdots, q_2^j\left(\boldsymbol{A_i}-q_1\left(\boldsymbol{A_i}\right)\right), \cdots, q_2^b\left(\boldsymbol{A_i}-q_1\left(\boldsymbol{A_i}\right)\right) \right).
$$
In the inverted file, vectors $\boldsymbol{A_i}$ are encoded using indices corresponding to the outputs of both $q_1(\boldsymbol{A_i})$ and $q_2(\boldsymbol{A_i} - q_1(\boldsymbol{A_i}))$. This two-level quantization approach strikes a balance between improving efficiency and capturing detailed information in the indexing process. 

The final stage, \emph{efficient similarity search}, navigates the inverted index to locate the $r$ nearest neighbors of a query vector. Given a query vector $\boldsymbol{x}$ and the inverted index built on database vectors $\{\boldsymbol{A_1}, \boldsymbol{A_2}, \cdots, \boldsymbol{A_N}\}$, FAISS first compares the distance between $\boldsymbol{x}$ and the centroids from the coarse quantizer $q_1$ to identify the clusters containing potential neighbors. 
$$
\mathcal{L}_{\mathrm{IVF}}= \underset{ \boldsymbol{c_i} \in \mathcal{C}_1}{\tau\text{-}\operatorname{argmin}}
\|\boldsymbol{x}-\boldsymbol{c_i}\|,
$$
where the multi-probe parameter $\tau$ is the number of coarse-level centroids considered during the search. Subsequently, FAISS scans the corresponding inverted lists of all the centroids in $\mathcal{L}_{\mathrm{IVF}}$. It computes the distance between subvectors $(\boldsymbol{x}-q_1(\boldsymbol{x}))^j$ and $\boldsymbol{c_i^j}$ for each subquantizer $q_2^j$ of the fine quantizer $q_2$, and then sums these distances to estimate the total distance from $\boldsymbol{x}$ to each vector in the scanned lists.
Ultimately, FAISS selects the $r$ nearest neighbors based on these estimated distances and returns their indices,
$$
\underset{i=1:N \text { s.t. } q_1\left(\boldsymbol{A_i}\right) \in \mathcal{L}_{\mathrm{IVF}}}{r\text{-}{\operatorname{argmin}}}\sum_{j=1}^b\left\|(\boldsymbol{x}-q_1(\boldsymbol{x}))^j-q_2^j(\boldsymbol{A_i}-q_1(\boldsymbol{A_i}))\right\|.
$$

\section{Implementation to Cyber Risk Literature: CyLit}\label{sec:cylit}
The amount of literature on cyber risk is increasing daily, due to growing awareness of cyber risk and cyber security. Cyber risk is a multifaceted issue that can be analyzed through various lenses, including analyzing monetary losses and legal consequences from cyber incidents and exploring ways to enhance cyber security. \citet{eling_cyber_2020} highlight the diverse range of topics covered in cyber risk literature and identifies ten categories of academic papers in this field based on the disciplines involved, such as management, economics, and telecommunications. 

\begin{figure}[h!]
    \centering  
    \includegraphics[width= 1.0\linewidth]{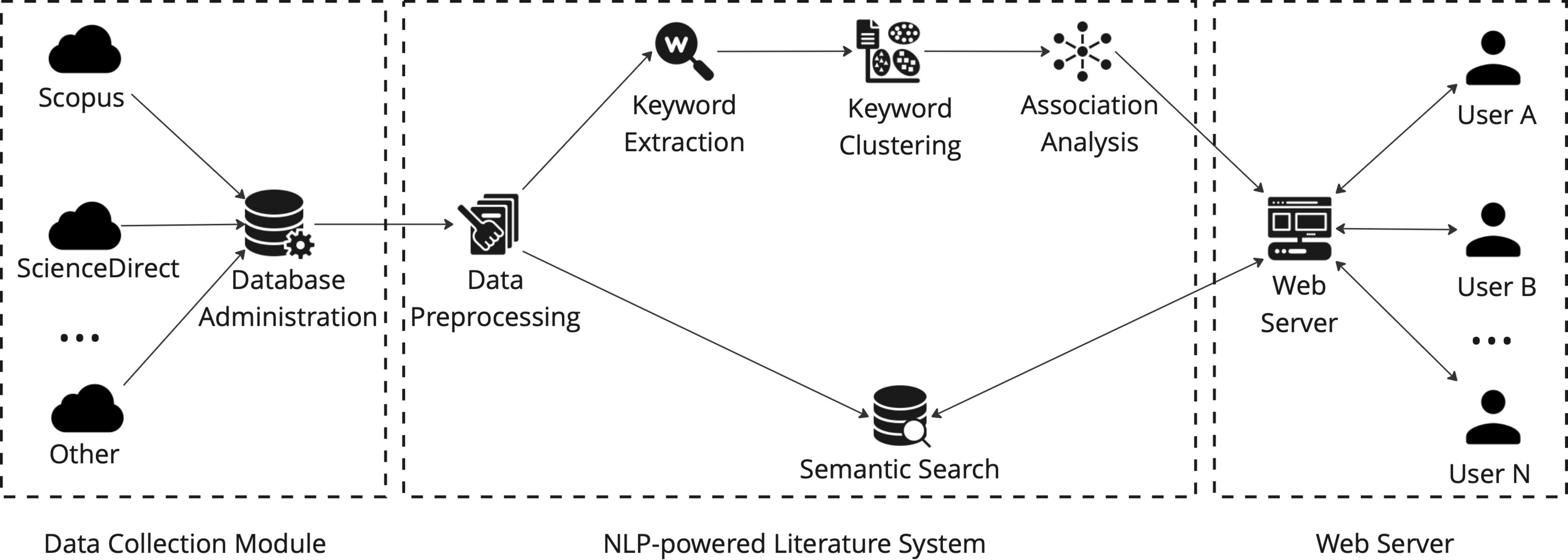}
    \caption{CyLit system structure illustration}
    \label{fig:arch}
\end{figure}

Many survey papers have been published on cyber risk; see \citet{berman_survey_2019, sardi_cyber_2020, aziz_systematic_2020, eling_cyber_2020}. However, as discussed in Section \ref{sec:intro}, current survey papers are limited by their static nature and reliance on manual review processes. For instance, \citet{eling_cyber_2020} offer a comprehensive overview of the cyber-related literature. Nevertheless, this survey paper covers only 217 papers, which is a small proportion of all cyber-related papers. In contrast, a query of the Scopus database as of February 25, 2023, yielded approximately 30,000 papers related to cyber risk, with the count continuing to rise. Additionally, as the survey was conducted in March 2020, it provides only a static snapshot of the cyber risk literature. Considering the dynamic nature of cyber-related issues due to rapid advancements in Internet technology, it is hypothesized that the current areas of concern may differ significantly from those two years ago. For example, ransomware attacks were once considered a major threat to companies, but the emergence of ransomware protection solutions offered by cloud service providers has greatly alleviated this problem, see \citet{renato_losio_cloud_nodate}.
To demonstrate the effectiveness of our approach described in Section \ref{sec:method} and facilitate cyber risk research, we have developed CyLit (see \citet{quan_cylit_2023}), an NLP-powered repository and search tool for cyber risk literature. Additionally, to enhance its utility, we have incorporated a data collection module and a web server, together with the NLP-powered literature system described in Section \ref{sec:method}. The structure of CyLit is shown in Figure \ref{fig:arch}.
\subsection{Data Collection}
Generally, two approaches can be used to obtain papers for a repository in a particular area of research. One approach scrapes web content, and the other utilizes Application Programming Interfaces (APIs) provided by literature databases. Using the former approach, it is necessary to identify a set of sites that consistently publish papers and allow automated tools to collect their content. While many resources fulfill the first criterion, few permit scrapings, making it a less viable option. This approach may be reserved for future research when the diversity of sources is the primary objective. The latter approach, which involves the use of literature database APIs, is preferred. This approach ensures that a vast collection of academic papers can be obtained, thereby facilitating the rapid expansion of the repository during its initial development stage.

In the current version of CyLit, the data collection module is mainly focused on collecting cyber risk literature from 
Scopus using 
its API due to its large volume of metadata, including title, abstract, and keywords, associated cyber-related articles. For future studies, other literature sources, such as Wiley and Crossref, can be included in the resources pool for data collection. 
The case-insensitive search query sent to Scopus
is as follows. 
\begin{verbatim}
    "Cybersecurity" OR "Cyberrisk" 
    OR "Cyber security" OR "Cyber risk"
    OR "Cyber literature" OR "Cyber insurance".
\end{verbatim}
The system retrieves and archives information on academic papers that match the query in their title, abstract, or keyword lists. Each collected paper's information includes unique identifiers, such as its identifier in Scopus and the Digital Object Identifier (DOI), along with literature metadata, such as paper information, author information, and publication information. Table \ref{tab:metadata} presents the selected metadata, including the paper's title, type, authors' names, abstract, author-provided keywords, and publication date. A comprehensive list of attributes is outlined in Appendix \ref{appendix_sec:article}. To comply with Elservier's licensing policy that the full text of academic papers cannot be displayed\footnote{https://www.elsevier.com/about/policies-and-standards/text-and-data-mining/license}, and as aforementioned, because concise and condensed information, such as titles and abstracts, is much more effective for our NLP models than the full-text data, we did not retrieve full-text articles from Scopus. 
However, the system retains the DOI and link to the publisher's site for each paper, which is made available to users who require access to the complete text. 
\begin{table}[htb]
    \centering
    \begin{tabular}{|l|l|}
         \hline
         \rowcolor{yellow}
        Metadata ID & Description\\
         \hline
            title & The title of the paper \\
            subtypeDescription & Type of paper  \\
            authorNames & Names of the authors \\
            description & The abstract of the paper \\
            authKeywords & Author-provided keywords \\
            coverDate & Publishing date \\
         \hline
    \end{tabular}
    \caption{Selected metadata in the paper}
    \label{tab:metadata}
\end{table}
 
To maintain consistency in our repository, the data collection module conducts basic data processing. This includes renaming or changing certain attributes of the newly collected papers to align with the format of existing data. In addition, the data collection module checks for duplicates. If a paper already exists in our repository, it discards the duplicate that is newly collected. This duplication check is especially useful for monthly data collection. To maintain the relevance and timeliness of the repository, we utilize cron jobs, a Linux utility that schedules job execution, to automate the fetching and processing of papers. To prevent repeated scanning of external sources, such as Scopus, for just a few hundred new records each time, the data collection unit sorts the results by publication date. Once the duplication check shows that the collected information starts to overlap with the existing data in our database, the data collection process terminates to avoid collecting duplicate information.

\subsection{NLP-powered Literature System}
We use the \texttt{authKeywords} provided in the metadata and followed the preprocessing procedures outlined in Section \ref{sec:preprocessing}. Subsequently, we have a comprehensive and representative keyword library that consists of 38,043 keywords and key phrases related to cyber risk literature. To generate word embeddings, we utilize the \texttt{all-MiniLM-L6-v2} model, which is a pre-trained sentence-transformer model based on the BERT architecture. This model maps the keywords and abstracts to a 384-dimensional vector space. Following the keyword extraction pipeline detailed in Section \ref{sec:keyword_extraction_and_clustering}, each paper has a set of keywords. We then apply K-means clustering, as discussed in Section \ref{sec:keyword_cluster}, to all the keyword and keyphrase embeddings. This process yielded 30 keyword clusters.
\begin{figure}[h!]
  \centering
  \includegraphics[width= 0.6\linewidth]{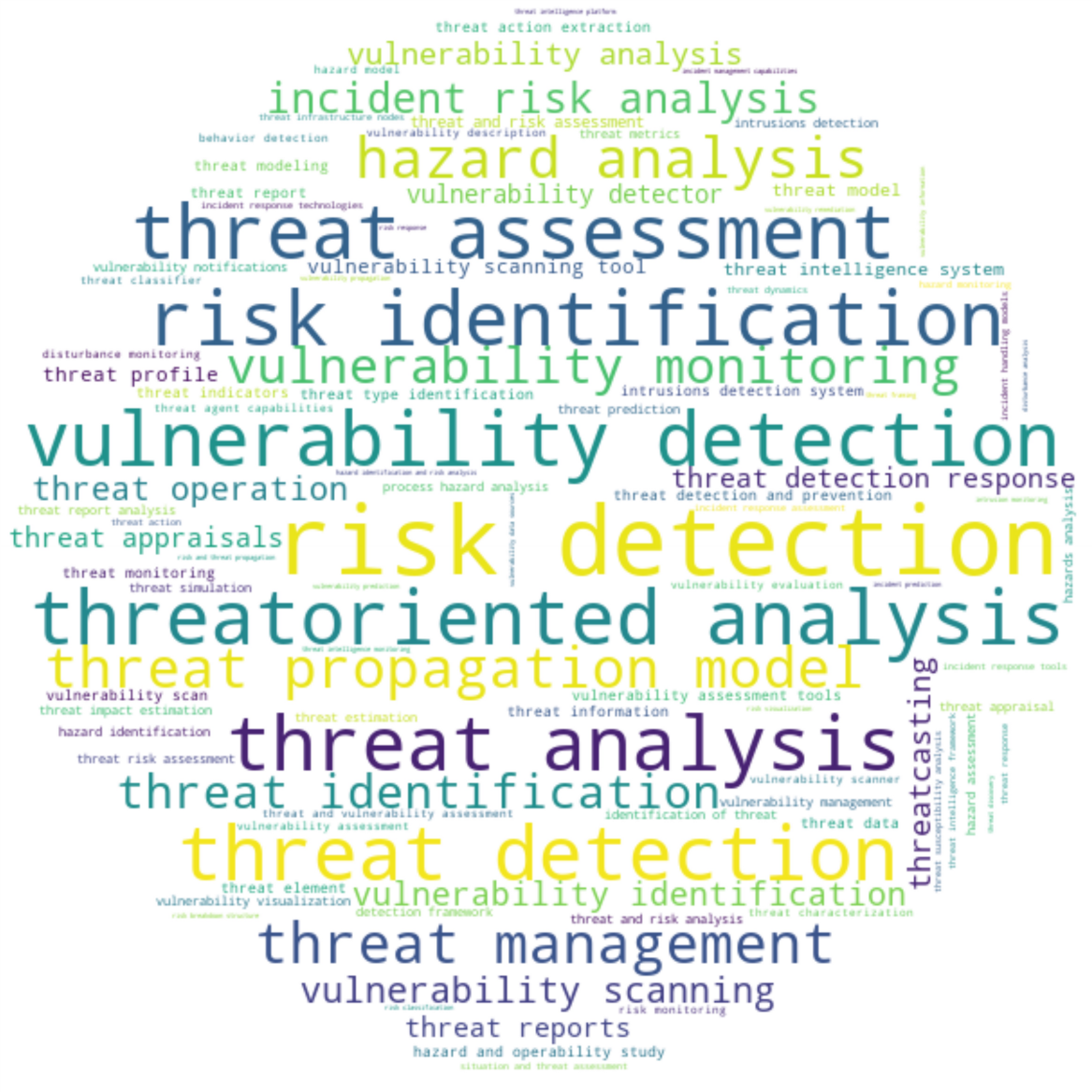}
  \caption{Word cloud for keyword cluster $C_1$ (Detection)}
  \label{fig:wordcloud_cluster_1}
\end{figure}

For visual representation, we generate word clouds for each keyword cluster. Each word in the cloud was sized according to its distance from the centroid of its cluster. Words closer to the centroid appeared larger, whereas those farther away appeared smaller. To ensure a fair comparison across clusters, we standardize the distances within each cluster based on the maximum distance observed in the cluster. Figure \ref{fig:wordcloud_cluster_1} shows the word cloud for keyword cluster $C_1$, centered around the keyword phrase ``risk detection''. This cluster mainly focuses on identifying risks and threats, as demonstrated by the presence of keywords such as ``risk identification'', ``vulnerability detection'', ``threat detection'', ``threat analysis'', and ``threat assessment'' close to its centroid. Therefore, we name this cluster ``detection''. For more information on keyword clusters, including their names, numbers of keywords, and numbers of associated papers, see Appendix \ref{appendix_sec:cluster}. Leveraging these keyword clusters facilitates the identification of cyber risk-related topics and the subsequent grouping of literature, thereby enhancing the efficiency of locating related information.

Analysis of the generated keyword clusters reveals a frequent co-occurrence of specific topics in papers, suggesting a potential association among these topics. For example,  the search for ``security issues in cyber-physical systems'' in CyLit has yielded a number of papers, such as \citet{agrawal_security_2022, dsouza_security_2019, bou-harb_brief_2016}, many of which contain keywords that fall under both $C_{8}$ (System Security) and $C_{15}$ (Cyberphysical Devices). To investigate cross-topic research activity in the field of cyber risk, an association analysis was performed using the Apriori algorithm outlined in Section \ref{sec:association}. The Apriori analysis results, as presented in Table \ref{tab:association}, were filtered based on various criteria, including $\text{support} \geq 0.05$, $\text{confidence} \geq 0.5$, and $\text{lift} \geq 1.5$, to identify association rules.

\begin{figure}[h!]
  \centering
  \includegraphics[width= 0.95\linewidth]{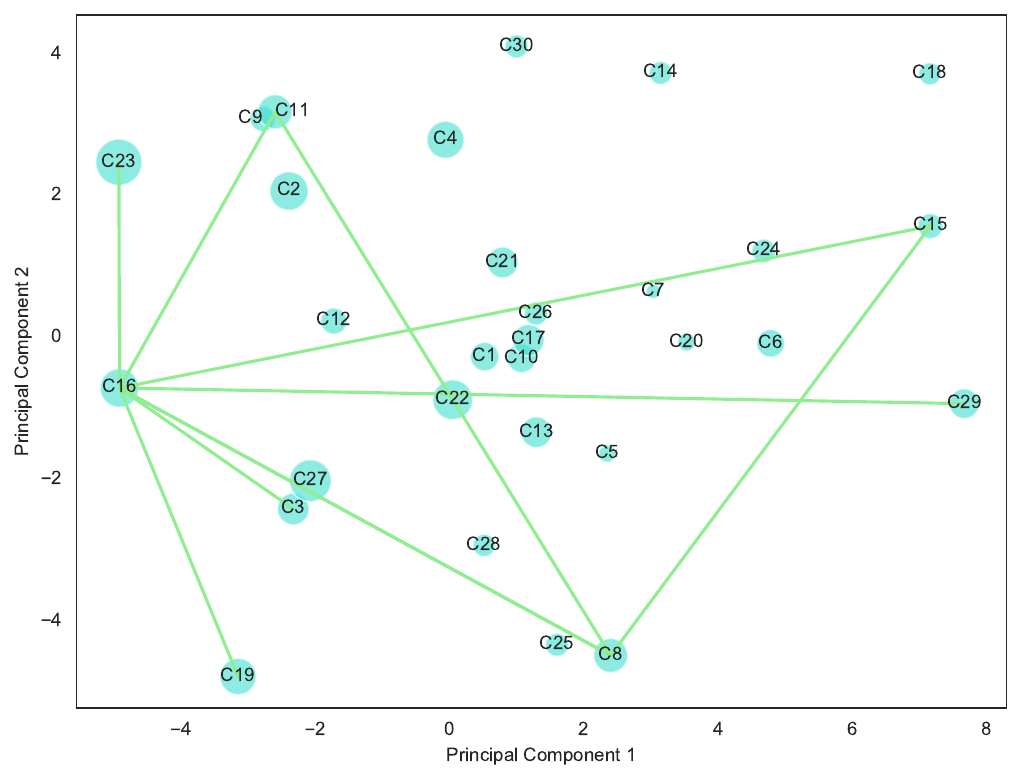}
  \caption{PCA projection of cluster centroids  with associations among clusters}
  \label{fig:network}
\end{figure}

\begin{table}[!ht]
\centering
\begin{tabular}{ L{.14\textwidth} L{.14\textwidth} 
L{.13\textwidth} L{.13\textwidth} L{.09\textwidth} L{.12\textwidth} L{.06\textwidth}}
\toprule
antecedents & consequents & antecedent support & consequent support & support & confidence & lift \\
\midrule
$C_{3}$ & $C_{16}$ & 0.230 & 0.352 & 0.132 & 0.573 & 1.630 \\
($C_{3}$, $C_{8}$)   & $C_{16}$ & 0.096 & 0.352 & 0.065 & 0.674 & 1.915 \\
($C_{3}$, $C_{11}$)  & $C_{16}$ & 0.068 & 0.352 & 0.051 & 0.745 & 2.119 \\
($C_{3}$, $C_{19}$)  & $C_{16}$ & 0.072 & 0.352 & 0.051 & 0.705 & 2.004 \\
($C_{3}$, $C_{29}$)  & $C_{16}$ & 0.135 & 0.352 & 0.073 & 0.540 & 1.535 \\
($C_{11}$, $C_{8}$)  & $C_{16}$ & 0.084 & 0.352 & 0.055 & 0.657 & 1.867 \\
($C_{11}$, $C_{16}$) & $C_{8}$  & 0.109 & 0.327 & 0.055 & 0.502 & 1.539 \\
($C_{15}$, $C_{8}$)  & $C_{16}$ & 0.097 & 0.352 & 0.058 & 0.597 & 1.698 \\
($C_{15}$, $C_{16}$) & $C_{8}$  & 0.101 & 0.327 & 0.058 & 0.576 & 1.765 \\
($C_{19}$, $C_{8}$)  & $C_{16}$ & 0.084 & 0.352 & 0.050 & 0.598 & 1.699 \\
($C_{23}$, $C_{8}$)  & $C_{16}$ & 0.085 & 0.352 & 0.051 & 0.594 & 1.688 \\ 
\bottomrule
\end{tabular}
\caption{Association rules from Apriori analysis}
\label{tab:association}
\end{table}
Figure \ref{fig:network} displays the semantic relationships and associations among the keyword clusters derived from the Apriori association analysis. Using Principal Component Analysis (PCA), we project the centroids of keyword clusters onto a two-dimensional plane. The node size is indicative of the number of keywords in each cluster, whereas the edges connecting the nodes represent the associations between the clusters, as presented in Table \ref{tab:association}. The proximity of the nodes (centroids of keyword clusters) indicates their semantic similarity. For instance, clusters $C_{3}$ (Cyber System Management) and $C_{27}$ (Assessment) are close to each other, just as $C_{17}$ (Electronic Control) and $C_{26}$ (Power System) are. Interestingly, the association rules from the Apriori association analysis reveal connections among clusters that are not always semantically close. For instance, the antecedents $C_{15}$ (Cyberphysical Devices) and $C_{16}$ (Miscellaneous Terms) are associated with the consequent $C_{8}$ (System Security) with a confidence of 0.576. This confidence indicates the probability of keywords from $C_{8}$ cooccurring with those from $C_{15}$ and $C_{16}$ in a paper. Additionally, the lift of 1.765 indicates a high positive correlation between the occurrence of keywords from $C_{15}$ and $C_{16}$ and the occurrence of keywords from $C_{8}$.

Effective storage and retrieval of data constitute a crucial aspect of CyLit. To facilitate this, a MongoDB-powered database system and an indexing system using FAISS are implemented in tandem, working collaboratively to achieve the desired objectives. Once the paper information is gathered from literature sources and supplemented with NLP-generated information, including extracted keywords and assigned keyword clusters, it is incorporated into the MongoDB database for storage. The database also accommodates specific queries, such as retrieving paper information based on publication type. Queries expressed in a semantic manner, such as ``the cost of a data breach'', may lack clarity for the database system, and thus are interpreted by the FAISS indexing system as detailed in Section \ref{sec:faiss}. Following the collection of a paper's abstract, its sentence embedding vector, generated by the NLP unit, is integrated into the indexing system. During a semantic search, FAISS performs a similarity comparison between the search query and the sentence embeddings in the system, returning the most relevant results.

\subsection{Web Server}
A web application\footnote{\url{https:/cylit.math.illinois.edu}} is built to make the repository accessible from any device anywhere. We take a standard web development approach to create the backend and frontend of the application. The backend is realized in Python using the Django framework on the basis of all the NLP models being implemented in Python. For the frontend, we swap out the Jinja\footnote{\url{https://jinja.palletsprojects.com/en/3.0.x/}} template, which is built in the Django framework, and replaced it with React\footnote{\url{https://react.dev/}}, which is a frontend library that has better support and scalability. The functionality of the web application includes paper lookup via filtering and semantic search, as well as some visualizations that provide an overview of the cyber risk literature from different perspectives. To be more specific, we can visually depict the distribution of papers per year within each keyword cluster. This visualization assists users in discerning trends across different keyword clusters. For instance, a first-year Ph.D. student intrigued by cyber risk can track the research trajectory over the years to identify compelling research topics. Furthermore, when users input specific queries in the search function, extending beyond retrieving papers related to the queries in cyber risk, the website can showcase the number of literature entries associated with the query over the years and present the corresponding keyword clusters. This integrated approach enhances the efficiency of the search function, amalgamating data visualization tools to unveil research trends and summary statistics related to the specified research topic.

\section{Human Review Compared to CyLit}\label{sec:survey}

In this section, we examine the validity of our clustering result by comparing it to some existing categorizations of cyber literature proposed in other scholarly works. Specifically, we refer to the results of the following three survey papers: 
\begin{itemize}
    \item \citet{berman_survey_2019} survey papers that use deep learning methods for cybersecurity tasks. Depending on the purpose of the application of deep learning, the authors put the literature in this field into 12 categories, among which some are niche areas, and three major categories include malware classification, malware detection, and intrusion detection. 
    \item \citet{sardi_cyber_2020} focus on cybersecurity issues in the healthcare industry. According to the origins of how those issues arise, the authors assign the relevant papers into three groups, including actions of people, systems and technology failures, and failed internal processes.
    \item \citet{aziz_systematic_2020} look at the challenges associated with cyber insurance and classifies related papers based on the key processes they focus on in insurance practice. The categories are organization eligibility, contract design, insured self-reporting, cyber insurance awareness, and the cost-benefit aspect.
\end{itemize}

Although these three papers have all created clusters for cyber-related papers, they clearly focus on different aspects of cyber risk, \textit{i.e.}, technical solutions to cybersecurity, cybersecurity issues in healthcare, and cyber insurance. Furthermore, they categorize papers based on various criteria, \textit{i.e.}, security issues to address, origins of security issues, and components in an insurance workflow, respectively. Even for papers on the same aspect of cyber risk, researchers can view them from different angles. For example, apart from \citet{aziz_systematic_2020}, another notable survey paper on cyber risk and insurance is \citet{eling_cyber_2020}, which groups research papers based on their fields of study, such as business research and quantitative or actuarial research. These discrepancies in how cyber-related papers are clustered illustrate that the task of literature categorization allows for great subjectivity and flexibility. Therefore, a comparison between our clustering result and those in the existing literature is conducted in a qualitative manner. Three questions are looked at in the comparison. Firstly, with respect to papers considered to be in the same group by other researchers, are they also in the same cluster according to our clustering result? Secondly, for papers that are considered to belong to different groups, are they also assigned to different clusters according to our result? Lastly, how can the differences be accounted for if our clustering result deviates from the clusters created by other researchers? 

Hereafter, we shall refer to the clusters created by other researchers as \textit{reference clusters} and the clusters developed in this study as \textit{CyLit clusters}. To make a comparison, we choose 36 papers listed in the three aforementioned survey papers, such that all reference clusters are of similar sizes. Then, we identify the CyLit clusters to which each of the same set of papers belongs. 

\begin{figure}
    \centering
    \includegraphics[width = \textwidth]{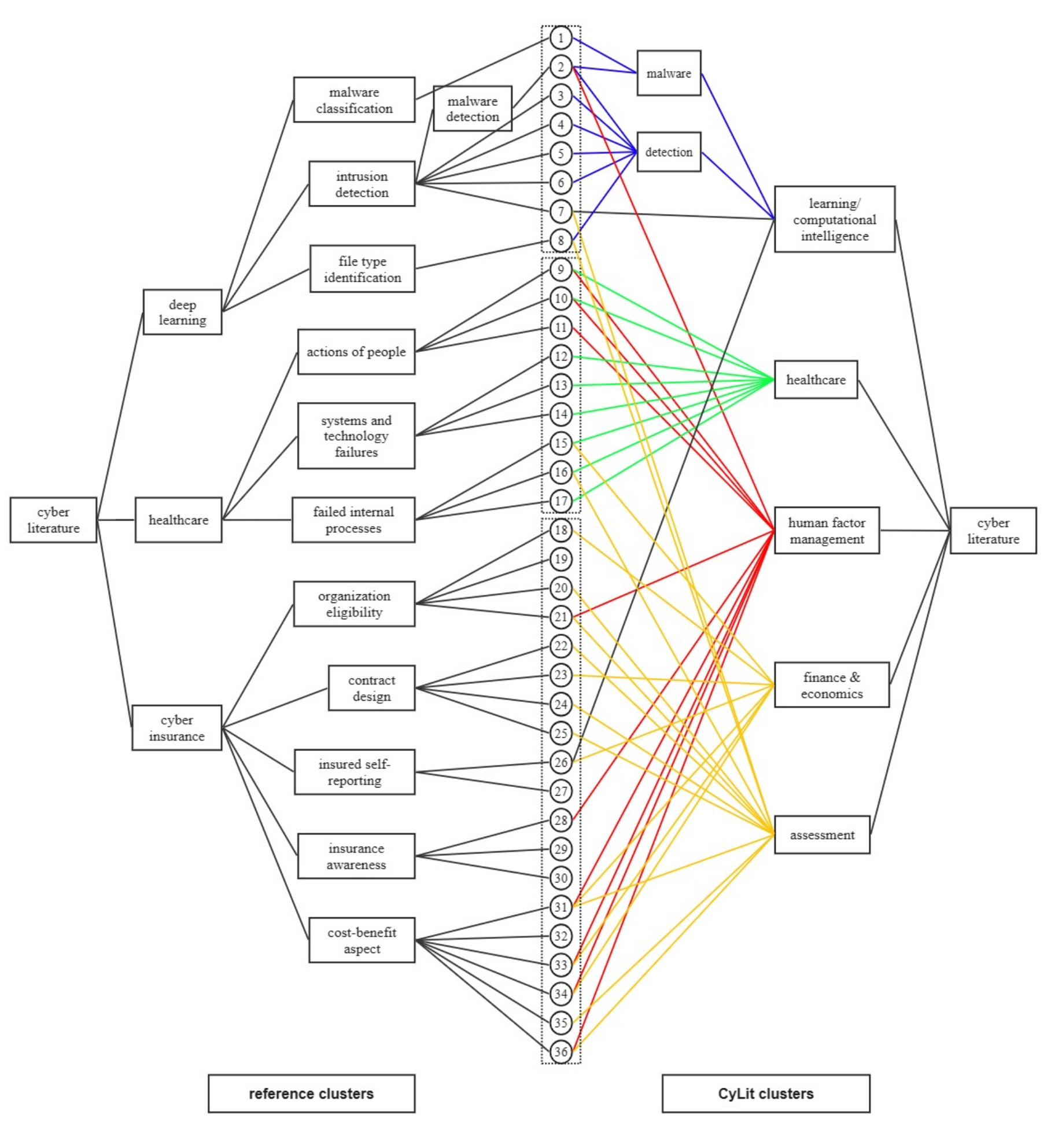}
    \caption{Comparison between reference clusters and CyLit clusters}
    \label{fig:cluster_compare}
\end{figure}

\subsection{Comparisons between Human Judgment and CyLit Clusters}
Three key observations made from the comparisons between reference and CyLit clusters are presented as follows. 
\subsubsection*{Cyber risk being interdisciplinary and multifaceted}
Reference clusters are constructed hierarchically, as presented in Figure \ref{fig:cluster_compare}, where each paper is exclusively associated with a single cluster. In contrast, the relationship between papers and CyLit clusters is many-to-many since papers are assigned based on multiple keywords. This is the first main difference between our clustering results and how existing works categorize cyber-related papers, which typically focus on a singular aspect of each paper to emphasize the theme of the survey. For example, Paper 11, \citet{priestman_phishing_2019}, is placed in the subcluster named ``actions of people" in \citet{sardi_cyber_2020}. Although this work focuses primarily on how healthcare organizations should counter the threat of phishing, some security measures involving human factors proposed in this paper, such as cyber security awareness training, are also applicable to many other organizations because phishing is a problem universally faced by organizations of all types. Therefore, in our clustering result, it is not surprising that \citet{priestman_phishing_2019} is assigned to both the healthcare cluster and the human factor management cluster, and the latter is not exclusive to healthcare papers but also includes cyber insurance papers that mention how organizations, in general, should manage their people in preparation for threats such as phishing. Many other papers in the sample of 36 papers are also assigned to multiple clusters, such as Paper 11 (see lines colored red in Figure \ref{fig:cluster_compare}). Because cyber-related papers are usually interdisciplinary and multifaceted, assigning a paper to multiple clusters can be more descriptive and accurate than assigning a single cluster.

\subsubsection*{Reliance of categorization tasks on domain knowledge}
Domain knowledge significantly influences the categorization methodologies employed by other researchers. For example, in \citet{aziz_systematic_2020}, cyber insurance papers are grouped according to some key components in the insurance business, such as underwriting, pricing, claim management, etc. To some extent, with the domain knowledge of the insurance business, the study conducted by \citet{aziz_systematic_2020} closely resembles a supervised classification task with known labels rather than an unsupervised clustering task. Without such domain knowledge, CyLit clusters, such as human factor management, finance \& economics, and assessment (see lines colored yellow in Figure \ref{fig:cluster_compare}), are created in a less systematic manner and lack a top-down structure, wherein lower-level clusters are compartments of an upper-level cluster. In this regard, the reference clusters provide more insights into the cyber literature related to specific disciplines than CyLit clusters. 

\subsubsection*{Non-uniform distribution of papers across disciplines}
The chosen survey papers represent three distinct disciplines, referred to as \textit{cybersecurity}, \textit{healthcare}, and \textit{cyber insurance}. As the $36$ papers for the comparison in this section do not constitute as a random sample of the cyber risk literature, their distribution across different disciplines is not proportional to the actual population distribution. Readers may refer to \citet{eling_cyber_2020} for insights into the volumes of cyber risk literature in different disciplines. Nevertheless, our results offer insights into the clustering perspective on the development of scholarly works related to cyber risk. 

Notably, despite the many-to-many relationship between papers and CyLit clusters, some hierarchical structure is still preserved for papers in the cybersecurity field. Specifically, all papers in the detection and malware clusters are also in the learning/computational intelligence cluster, suggesting that, with keywords extracted from or provided by the existing cyber risk literature, the K-means algorithm can distinguish between niche areas of cybersecurity (see the lines colored blue in Figure \ref{fig:cluster_compare}). For example, in \citet{berman_survey_2019}, Paper 1 is categorized as related to malware classification and different from any intrusion detection papers. Paper 2 is also about malware but is related to detection instead of classification. The commonality of the two papers is that they are both related to the application of deep learning. The subtle difference and the common trait of these two papers are captured by CyLit clusters, \textit{i.e.}, for CyLit clusters, Paper 1 is assigned to the malware cluster, and Paper 2 is assigned to both the malware and the detection clusters. Both papers are in the learning/computational intelligence cluster. Enabled by CyLit clusters, the high-resolution view of cybersecurity topics results from a large volume of literature focusing on different aspects of cybersecurity. In contrast, in terms of papers related to healthcare, although \citet{sardi_cyber_2020} create granular clusters, such as systems and technology failures and failed internal processes, CyLit clusters fail to recognize the difference among them in general due to relatively fewer papers in these two disciplines (see the lines colored green in Figure \ref{fig:cluster_compare}). 

\subsection{Remarks on the Difference between Human and Machine Approaches}
In sum, regarding the three questions raised earlier, CyLit can put papers into broad groups similar to the scopes of the existing survey papers that focus on specific disciplines. However, depending on the volume of literature in the cyber-related discipline, CyLit clusters may or may not capture niche topics within each broad group. CyLit clusters align well with the reference clusters created by other researchers for disciplines with a large volume of literature. Otherwise, discrepancies between CyLit and reference clusters may occur. In addition, domain knowledge is used in conventional survey papers to create clusters for literature, whereas without that domain knowledge, CyLit produces less structured clusters. Moreover, compared to conventional survey papers, CyLit has the advantage of providing a comprehensive view of various topics covered by a paper and discovering the interdisciplinary relations among papers by assigning each paper to multiple clusters. 

Lastly, it should be highlighted that manually surveying the literature requires researchers to devote a tremendous amount of time and effort, and progress may still fall behind the growth of the paper volume, especially so in rapidly expanding research areas such as cyber risk. In that case, a machine-based literature categorization and search tool like CyLit can serve as a complementary information source, and its efficiency and scalability provide researchers with a comprehensive and up-to-date overview of the landscape of research in a certain field, which is usually missing in conventional survey papers.

\section{Large Language Models Compared to CyLit}\label{sec:llm}

Through the comparisons between human and machine approaches, the previous section highlights their distinct strengths and advocates the integration of CyLit and manual review as a complete workflow of literature review. This workflow takes advantage of both the scalability and efficiency of CyLit and scholars' domain knowledge and in-depth analysis. 

Because ChatGPT, a prominent example of Large Language Models (LLM), generates requested information seamlessly and is backed by ever-updating knowledge sources, whether or not it can be a feasible replacement for the proposed workflow is a natural question. Therefore, for the last piece of this study, the performance of ChatGPT in this respect shall be examined closely.

\subsection{Brief Overview of LLM} 

LLM are a class of machine learning models designed to understand and generate human-like text based on vast amounts of data. These models have revolutionized the field of NLP, offering unprecedented capabilities in text analysis, generation, and comprehension. The work by \citet{vaswani2017attention} represents a notable milestone in the progression of NLP, as it introduces the transformer architecture, including models like BERT, as discussed earlier in Section \ref{sec:method}. Generative Pre-Trained Transformers (GPT),  introduced by \citet{radford_improving_2018}, is another transformer-based model that uses a decoder-only architecture. Unlike BERT's bidirectional training, GPT adopts an autoregressive language modeling approach, operating in a unidirectional fashion from left to right to predict the next word in a sequence based on the preceding words. Every word can only attend to previous words in the sequence, aligning with the natural process of language generation where each word depends on the ones that come before it. The design enables GPT to excel in text-generation tasks. Based on this foundation, GPT-2, introduced by \citet{radford_language_2019}, marked a significant leap in language generation capabilities. It is trained on a much larger dataset and designed to generate more coherent and longer passages of text. In 2020, \citet{brown_language_2020} reveal GPT-3, which dramatically scales up the model's size and complexity, boasting 175 billion parameters. Following GPT-3, the GPT series continued to evolve with GPT-3.5 and GPT-4\footnote{\url{https://cdn.openai.com/papers/gpt-4.pdf}}, each iteration bringing enhancements in language comprehension, contextual understanding, and text generation capabilities. These models represent the forefront of Generative AI, demonstrating capabilities ranging from creative writing to complex problem-solving and have been instrumental in developing applications such as ChatGPT\footnote{\url{https://chat.openai.com}}, developed by OpenAI.

Derived from the GPT architecture, ChatGPT has been fine-tuned to generate human-like text in a conversational format based on the prompts it receives from the user. Its capabilities span a range of language-based tasks, encompassing answering queries, providing explanations, crafting creative content, language translation, and more. Notably, ChatGPT has attained top ranking\footnote{\url{https://huggingface.co/spaces/lmsys/chatbot-arena-leaderboard}} among various LLM showcased on platforms such as  Bard\footnote{\url{https://bard.google.com}} by Google, Claude\footnote{\url{https://claude.ai}} by Anthropic, and LLaMA\footnote{\url{https://ai.meta.com/llama}} by Meta.

\subsection{Comparisons between LLM and CyLit Clusters}
We undertake multiple experiments to evaluate the feasibility of employing LLM, particularly the latest version, ChatGPT-4, in automatically generating summaries and literature reviews, as opposed to the combination of CyLit and manual literature review\footnote{The experiments were conducted in January and February 2024.}. In this subsection, we outline the challenges in literature categorization and review and the inadequacies of ChatGPT in addressing them. For a comprehensive overview of the prompts and responses from ChatGPT utilized in the experiments, please refer to Appendix \ref{appendix_sec:chatgpt_experiments}.

\subsubsection*{Need for tailored approaches to literature categorization}

The experiment encompasses the 36 papers examined in Section \ref{sec:survey}, originating from three survey papers focusing on different aspects of cyber risk. These papers are selected to assess ChatGPT's ability to identify and summarize distinct topics within these papers, associate pertinent literature with each topic, and provide concise summaries for each identified cluster. 

Due to the limitations in processing the contents of a large number of PDF files and the constraints in the input text token capacity for ChatGPT, we compile an Excel file containing solely the titles and abstracts of the 36 papers. ChatGPT is then prompted to review these papers and categorize them into distinct groups based on their research topics. Furthermore, it is instructed to assign a descriptive name to each group, reflecting the common theme shared by the papers within it, and to provide the rationale behind its categorization. The experiment is repeated three times with identical inputs (data and prompt) to ascertain the consistency and validity of the outcomes. 

The analysis of ChatGPT's responses and its underlying methodologies shows that the model adopts a straightforward approach to processing the given texts. The initial step involves using TF-IDF method to vectorize the combined text of each paper's title and abstract. Following this, ChatGPT either employs K-means clustering or Latent Dirichlet Allocation (LDA), a topic modeling technique that assumes documents as mixtures over an underlying set of topics (see \citet{blei2003latent}).  
The variability in the employed methods accounts for the inconsistency in the results. In some instances, ChatGPT utilized K-means clustering, while in others, it applied LDA. Furthermore, the decision to use or omit a seed number for the random state in these algorithms varied with each experiment. This variability led to non-replicable and unstable outcomes. A fixed preset number of five topics, applied without adjustments based on the results, highlighted a limitation in ChatGPT's ability to dynamically tailor its approach to the dataset at hand.

These naive approaches resulted in overly broad categories and inaccurate characterizations of papers. For example, in the second trial of this experiment, one of the created clusters is named ``Risk Management and Cyber Risk in Various Sectors,'' which only delivers an obscure meaning. Note that although each of the CyLit clusters is also broad-ranging with a name such as ``healthcare'', each paper is assigned multiple labels, thus having more precise and specific characterizations. Also in this trial, Paper 30, which is about defense resource allocation from an insurance perspective (see \citet{lau_cybersecurity_2020}), is erroneously placed in the same group as other papers on ``Deep Learning and Anomaly Detection in Cybersecurity''. These results suggest that the categorization conducted by ChatGPT is subject to great arbitrariness, and specific tools such as CyLit are needed to generate more precise categorizations of papers. 

\subsubsection*{Necessity of sophisticated text processing and analysis}

In light of the unsatisfactory performance observed in the initial experiment, we undertake a follow-up experiment that incorporates more directed prompts for ChatGPT. Rather than allowing ChatGPT to independently categorize the 36 papers into distinct groups based on its interpretation of these research topics, we provide ChatGPT with predefined categories derived from human judgments stemming from the original survey paper: Deep Learning, Healthcare, and Cyber Insurance. The objective is to assess whether providing such guided information would enhance ChatGPT's categorization accuracy.

This approach generally produces improved results, particularly in the classification of papers within the Deep Learning category. Nevertheless, challenges persist in accurately categorizing articles related to Healthcare and Cyber Insurance, and some articles remain uncategorized.

The examination of ChatGPT's methodology reveals its straightforward, keyword-based approach to categorization. ChatGPT initially establishes a set of keywords associated with each topic, employing a somewhat opaque process. Subsequently, it implements a scoring function, wherein the relevancy score of a paper to a category is gauged by the frequency of the predetermined set of keywords appearing in the title and abstract of a paper. Papers are subsequently classified based on the highest relevancy score for a given topic. This approach hinges on the exact match of predefined keywords within the text, overlooking semantic meanings compared to our approach. 

Notably, no preprocessing is performed on the text, leading to some articles remaining uncategorized due to a lack of keyword matches. An exemplary instance highlighting the limitations of this approach is the paper \citep{akinsanya_towards_2020} titled ``Towards a maturity model for health-care cloud security (m2hcs)'', which is not categorized under Healthcare. This discrepancy stems from the usage of ``health-care'' instead of ``healthcare'' in paper's title and abstract, diverging from the ChatGPT's predetermined set of keywords for Healthcare, which included keywords such as ``healthcare'', ``medical'', ``patient data'', ``health sector'', and ``clinical''. Conversely, such nuances are addressed in our preprocessing steps within the CyLit system, ensuring successful identification and association with the relevant CyLit cluster. Furthermore, ChatGPT's approach fails to recognize the relationship between semantically similar words that may appear distinct. For instance, cyber warranties, which represent coverage offered by security providers in case of losses, bear significant similarities to insurance. However, the paper \citep{bushnell_cyber-warranties_2018} titled ``Cyber-warranties as a quality signal for information security products'' is not categorized under Cyber Insurance. This indicates that ChatGPT does not incorporate essential domain knowledge into the categorization process. It is worth mentioning that the two examplar papers are successfully associated in our CyLit system.
Utilizing a comparable keyword-based approach, our CyLit system showcases a more advanced and efficient methodology for literature review and categorization tasks, incorporating detailed preprocessing, keyword extraction, and clustering with semantic analysis.

In light of the limitations identified in ChatGPT's approach in previous experiments, we conduct a supplementary experiment aimed at guiding ChatGPT to adopt methodologies similar to those employed in our system, specifically focusing on keyword extraction and clustering. We prompt ChatGPT to perform keyword extraction using KeyBERT for each paper without author-provided keywords, and then proceed with keyword clustering as mentioned in Section \ref{sec:keyword_extraction_and_clustering}. For each identified keyword cluster, ChatGPT is instructed to list the keywords alongside the IDs and titles of the associated papers. However, this experiment encounters a significant obstacle: ChatGPT reports errors while attempting to execute the task due to the unavailability of the KeyBERT library in its operating environment. This limitation underscores a fundamental constraint of ChatGPT's current capabilities—it cannot execute code or directly interact with external software libraries, such as KeyBERT. Faced with this constraint, ChatGPT proposes an alternative approach utilizing TF-IDF for keyword extraction, demonstrating its adaptability but also highlighting the constraints of its operational environment. This experiment further illuminates the challenges associated with performing advanced computational tasks within ChatGPT's environment. Despite its potential for running Python code in a Jupyter-like setting, the platform's limited access to specialized libraries restricts its ability to perform sophisticated data processing tasks. The computational errors encountered across multiple experiments emphasize the ChatGPT's current boundaries in executing real-time computations or interfacing with a broader range of computational tools and methodologies.        

\subsubsection*{Critical reviews' dependence on knowledge outside the text}

As we propose using CyLit to help human conduct critical reviews, a natural question arises: can ChatGPT directly do critical reviews without human intervention? To examine ChatGPT's ability to conduct critical reviews of academic papers, it is prompted to read the full text of three papers and give a summary and in-depth analysis of various attributes of the paper, including methodology, findings, and contributions. Note that this experiment is to test the possibility of ChatGPT replacing human reviewers. It does not constitute a direct comparison between LLM and CyLit regarding critical reviews.

The summaries of articles given by ChatGPT are overall satisfactory. Key information has been effectively extracted, summarized, and highlighted in the response. Some points brought up by the authors are reorganized and potentially presented in a more efficient way. For example, in \citet{yousefi-azar_autoencoder-based_2017}, two of the four contributions given by the authors are:  

\begin{quote}
    ``Our scheme uses almost the minimum number of features compared to other state-of-the-art algorithms. This makes the model to be more effective for real time protection.
    
    In addition to the limited number of original features, the proposed scheme generates a small set of latent features. The resulting rich and small latent representation makes it practical for it to be implemented in small devices such as the Internet of Things.''
\end{quote}
Both paragraphs essentially state that the proposed scheme is lightweight and can be used in various scenarios that allow for limited computing time and/or resources. The response given by ChatGPT combined these two points and stated the following:
\begin{quote}
    ``The research outlines the practical implications of the proposed scheme, noting its efficiency in using a minimal number of features and its applicability to real-time protection and implementation in resource-constrained devices such as IoT devices.''
\end{quote}

However, this ability to accurately summarize the text also makes ChatGPT a less competent critic. When ChatGPT commented on the contributions of a piece, in most cases, it simply rephrased and summarized the contributions listed by the authors of the article rather than taking a global view of the literature to objectively assess the novelty and value created by the reviewed article. This may give users and readers a wrong impression of the importance of the paper, and in this aspect, human input still cannot be replaced.

Similarly, when commenting on the limitations of a study, the response was usually based on the limitations mentioned by the authors without referring to the insufficiencies of the work compared to other literature in the corresponding field. For example, \citet{kessler_information_2020} has identified that the limitations of their work include the lack of cross-sectional data for comparisons between different demographic groups. The comment on the limitation of this work given by ChatGPT simply reiterated this point. 

In addition, when asked to provide some critical perspectives on the paper, ChatGPT sometimes gave comments that were detached from the content of the paper. For example, regarding \citet{nguyen_cyberattack_2018}, the response given by ChatGPT included the following statement:

\begin{quote}
    ``Additionally, exploring the impact of different types of cyberattacks on the model's performance and how it adapts to new, previously unseen attack vectors could further validate its robustness and adaptability.''
\end{quote}
This point has been addressed by the authors of this paper by applying the proposed approach to the NSL-KDD dataset, which has 24 attack types in the training set and 38 attack types in the test set. Technically, the performance of the model has been tested on previously unseen attack vectors. Therefore, this comment made by ChatGPT does not offer much constructive value. This may suggest that it has difficulties in capturing information that is not explicitly conveyed.

\subsection{Remarks on Limitations of General-Purpose LLM and Potential Improvements}

The findings of these experiments reveal the current limitations of commercial off-the-shelf LLM for intricate tasks like categorizing papers into specific topics and conducting thorough literature reviews within niche research domains. One potential avenue for improvement involves fine-tuning open source LLM, which offers greater flexibility for domain-specific training and customization. For instance, leveraging an open-source model like LLaMA \citep{touvron2023llama, touvron2023llama2}, recognized for its excellence among open-source LLM, for downstream fine-tuning using domain-specific papers could bolster the LLM's capacity to comprehend and process specialized content within distinct research fields.

Nevertheless, this approach presents its own challenges. A significant obstacle is the text input limitation predetermined by the pre-trained LLM training. To mitigate this, a viable strategy is to convert the text into embeddings, thereby condensing the information into a format more manageable for the pre-trained LLM. Combining this approach with domain-specific embedding training holds promise for refining the model's performance in tasks such as academic literature review.

However, employing open-source LLM for domain-specific training necessitates substantial computational resources. Training and fine-tuning these models demand computational power and funding resources that may exceed the capabilities of smaller research groups like ours, and can pose a challenge even for most of academic scholars unless they collaborate with large tech companies with extensive resources. If such resources are available, training domain-specific LLM from scratch could address the aforementioned issues more effectively.

\section{Conclusion and Future Directions}\label{sec:conclusion} 
Interest in cyber risk research is on the rise, but with a growing volume of literature in this field and the interdisciplinary nature of this topic, it becomes difficult for researchers to find the information that is most helpful to their research questions. In this project, we build CyLit, an NLP-powered repository and search tool for cyber risk literature. The repository is self-updating, thus staying relevant to the latest topics in the field of cyber risk. NLP techniques enhance the capabilities of the CyLit system in several dimensions: heightened accuracy and efficiency through automated processes like categorizing relevant papers; precise interpretation and response to users' search queries, thereby saving users' time and reducing search errors; delivering richer insights by extracting valuable information from extensive unstructured text data on cyber-related topics; and offering improved summarization by generating concise summaries for cyber-related papers and identifying key themes and topics for users. Additionally, the repository is equipped with a web application, which makes querying the repository easy. All these features allow cyber risk researchers to locate the needed information efficiently.  

To demonstrate the performance of this tool, we compare its categorization results to categories in survey papers and with categories created by ChatGPT, the interface of an exemplary LLM. The comparison shows that CyLit provides unique insights complementary to the perspectives of human reviewers. While ChatGPT excels at many generative tasks, it does not provide tailored solutions to the domain-specific categorization problem. This limitation can potentially be overcome by fine-tuning the model downstream using domain-specific papers, and the integration of a fine-tuned open-source LLM into CyLit can be a future direction of research on living literature review.

We can extend this framework to encompass broader actuarial science research or concentrate on specific actuarial research domains characterized by multidisciplinarity, such as loss modeling, climate risk, etc. In the foreseeable future, actuarial science research is poised to expand, where researchers find it challenging to keep up with a multitude of publications and stay abreast of current research trends. Implementing an automated literature system will significantly save time for users and foster interdisciplinary research.

We recognize the importance of including user-centric evaluations in our research. While the current study has focused on the development and theoretical underpinnings of our methodologies, future work will expand to assess their practical efficacy in user-centric scenarios, for example, user studies and experiments, user experience analysis, and iterative design and improvement. By incorporating these user-focused research activities, we aim to ensure that our work is not only academically robust but also valuable and effective for end-users.   

\section*{Acknowledgements}
The authors are grateful to anonymous reviewers for their careful reading and insightful comments. Funding for this project is provided by the Campus Research Board, University of Illinois at Urbana-Champaign. This work is also supported by a General Insurance Research Committee (GIRC) Research Grant (2022) from the Society of Actuaries (SOA). Any opinions, findings, conclusions or recommendations expressed in this material are those of the authors and do not necessarily reflect the views of the SOA. 

\clearpage

\printbibliography

\clearpage

\begin{appendices}
\section{Mathematical Notations}
\label{appendix_sec:notation}

\begin{longtable}{@{}ll@{}}
\toprule
Symbol & Description \\
\midrule
\endfirsthead

\multicolumn{2}{r}{{Continued from previous page}} \\
\toprule
Symbol & Description \\
\midrule
\endhead

\midrule
\multicolumn{2}{r}{{Continued on next page}} \\
\endfoot

\bottomrule
\\
\caption{Summary of symbols and their descriptions} \label{tab:notation} \\
\endlastfoot

$D$ & The literature repository, the set of all the papers. \\
$d_i$    & The $i$-th paper in the literature repository $D$. \\
$A_i$   & The abstract of the $i$th paper. \\
$K_i$ & The set of keywords from the paper $d_i$. \\
$kw_{ij}$ & The $j$-th keywords in the set of keywords $K_i$. \\
$K$ & The consolidated keyword library, the set of all the keywords. $K =\bigcup_{i=1}^{N_1} K_{i}$. \\
$kw_{i}$ & The $i$-th keywords in the keyword library $K$. \\
$M$ & Number of keywords in the keyword library. $M=|K|$. \\
$N_1$  & Number of papers in the repository with author-provided keywords. \\
$N_2$  & Number of papers in the repository without author-provided keywords. \\
$N$  & Number of papers in the repository. $N = N_1 + N_2$. \\
\midrule
$W_i$ & The set of tokens extracted from $A_i$. \\
$w_{ij}$ & The $j$-th token extracted from $A_i$. \\
$\boldsymbol{w_{ij}}$ & The embedding of $w_{ij}$. \\
$\boldsymbol{A_i}$   & The embedding of the abstract $A_i$. \\
$\alpha$ & Diversity parameter in MMR. \\
$K_i^s$ & The set of selected keywords from $A_i$ by KeyBERT.\\
$w_{ij}^s$ & The $j$-th selected keywords in $K_i^s$.\\
$\boldsymbol{w_{ij}^s}$ & The embedding of $w_{ij}^s$. \\
$m$ & Number of selected keywords from $A_i$ by KeyBERT. $m=|K_i^s|$.\\
$W_i^c$ & The set of candidate keywords that are most relevant to the paper, based on\\
 & the highest cosine similarity values.  $\left|W_i^c \right| = 2m$ in Max Sum Distance algorithm.\\
$w_{ij}^c$ & The $j$-th candidate keywords in $W_i^c$. \\
$\boldsymbol{w_{ij}^c}$ & The embedding of $w_{ij}^c$. \\
\midrule
$\boldsymbol{kw_{i}}$ & The embedding of $kw_{i}$. \\
$k$ & The number of keyword clusters. \\
$C_j$ & The $j$-th keyword cluster.\\
$\boldsymbol{c_{j}}$ & The centroid of the $j$-th keyword cluster $C_j$.\\
$|C_j|$ & The number of keywords in the keyword cluster $C_j$.\\
\midrule
$R_s$  & The $s$-th association rule.\\
$LHS(R_s)$ & The antecedents of $LHS(R_s) = \{ C_\mathcal{U} \mid \mathcal{U} \subseteq  \{1, 2, \cdots, k \} \}$. \\
$RHS(R_s)$ & The consequents of $RHS(R_s) = \{ C_\mathcal{V} \mid \mathcal{V} \subseteq  \{1, 2, \cdots, k \} \}$. \\
$N_\mathcal{U}$ & The number of papers containing keywords from all the clusters in $LHS(R_s)$.\\
\midrule
$p$ & The number of dimensions for the embedding $A_i$.\\
$x$ & A query. \\
$\boldsymbol{x}$ & The embedding of the query $x$.\\
$r$ & The number of papers that are most semantically close to the query $x$, which\\
 & are returned as results by FAISS.\\
$\boldsymbol{y}$ & A p-dimensional vector. \\
$q$ & A quantizer function.\\
$\boldsymbol{c_i}$ & The $i$-th centroid from a quantizer.\\
$\mathcal{C}$ & A codebook that is a set of all the centroids from a quantizer.\\
$|\mathcal{C}|$ & The size of $\mathcal{C}$.\\
$\mathcal{I}$ & The index set of the codebook $\mathcal{C}$. \\
$b$ & The number of subvectors in product quantization.\\
$\boldsymbol{y}^j$ & The $j$-th subvector of $\boldsymbol{y}$. \\
$q^j$ & The quantizer for the $j$-th subvector in product quantization.\\
$q_1$ & The coarse quantizer.\\
$\mathcal{C}_1$ & The codebook of the coarse quantizer $q_1$.\\
$\left|\mathcal{C}_1\right|$ & The size of the codebook $\mathcal{C}_1$.\\
$q_2$ & The fine quantizer.\\
$q_2^j$ & The $j$-th subquantizer of the fine quantizer $q_2$.\\
$\mathcal{L}_{\mathrm{IVF}}$ & The list of coarse-level centroids that are semantically closest to the query.\\
$\tau$ & The multi-probe parameter, which is the number of coarse-level centroids\\
& considered during the search.\\

\midrule
$||\cdot||$ & Euclidean distance. \\
$\text{sim} \left( \cdot, \cdot \right)$ &  Cosine similarity.\\
\end{longtable}

\clearpage

\section{Paper Information Collected}
\label{appendix_sec:article}
\begin{table}[h]
\centering
\resizebox{!}{8.9cm}{
\begin{tabular}{@{}ll@{}}
\toprule
Column Name          & Column Definition                                \\ \midrule
affiliation\_city    & Affiliation city                                 \\
affiliation\_country & Affiliation country                              \\
affilname            & Affiliation name                                 \\
afid                 & Affiliation ID                                   \\
aggregationType      & Type of publication (Book, Journal, etc.)        \\
article\_number      & Paper number                                   \\
authkeywords         & Author provided keywords                         \\
author\_afids        & Author affiliations                              \\
author\_count        & Number of authors                                \\
author\_ids          & Author IDs                                       \\
author\_names        & Author names                                     \\
citedby\_count       & Number of times that this paper is cited       \\
coverDate            & Publication date                                 \\
coverDisplayDate     & Publication year                                 \\
creator              & Corresponding author                             \\
description          & Abstract                                         \\
doi                  & Digital Object Identifier                        \\
eIssn                & Electronic International Standard Serial Number  \\
eid                  & Scopus EID                                       \\
fund\_acr            & Sponsor acronym                                  \\
fund\_no             & Grant number                                     \\
fund\_sponsor        & Sponsor name                                     \\
identifier           & Scopus ID                                        \\
issn                 & International Standard Serial Number             \\
issueIdentifier      & Issue                                            \\
openaccess           & Open access flag (0 or 1)                        \\
openaccessFlag       & Open access flag (False or True)                 \\
pageRange            & Page range                                       \\
pii                  & Publisher Item Identifier                        \\
publicationName      & Publication name (Journal name, book name, etc.) \\
source\_id           & Scopus source ID                                 \\
subtype              & Subtype code                                     \\
subtypeDescription   & Subtype description (Review, article, etc.)      \\
title                & Title of paper                                 \\
url                  & Link to paper                                  \\
volume               & Volume                                           \\ \bottomrule
\end{tabular}}
\end{table}

\clearpage

\section{Keyword Clusters at a Glance}
\label{appendix_sec:cluster}

\begin{table}[h]
\centering
\begin{tabular}{@{}llll@{}}
\toprule
Cluster No. & Cluster Name                       & \# of Keywords & \# of Papers \\ \midrule
$C_1$                     & Detection                          & 1150               & 4866             \\
$C_2$                      & Cyber Terminology                  & 2000               & 5086             \\
$C_3$                      & Cyber System Management            & 1392               & 6588             \\
$C_4$                      & Cyber Regulations \& Public Policy & 1855               & 3966             \\
$C_5$                      & Healthcare                         & 495                & 971              \\
$C_6$                      & Computational Intelligence         & 1090               & 5709             \\
$C_7$                      & Mobile                             & 377                & 1075             \\
$C_8$                      & System Security                    & 1588               & 9372             \\
$C_9$                      & Misc. I                            & 944                & 2821             \\
$C_{10}$ & Finance \& Economics               & 1169               & 3114             \\
$C_{11}$ & Security Breach                    & 1597               & 6104             \\
$C_{12}$ & Misc. II                           & 1002               & 3038             \\
$C_{13}$ & Technology Management              & 1315               & 5203             \\
$C_{14}$ & Malware                            & 800                & 3289             \\
$C_{15}$ & Cyberphysical Devices              & 925                & 6813             \\
$C_{16}$ & Misc. III                          & 2015               & 10097            \\
$C_{17}$ & Electronic Control                 & 1365               & 5017             \\
$C_{18}$ & Cyber Crimes                       & 753                & 2711             \\
$C_{19}$ & System Resilience                  & 1787               & 6114             \\
$C_{20}$  & Cryptography                       & 526                & 1454             \\
$C_{21}$ & Algorithm                          & 1322               & 2536             \\
$C_{22}$ & Data Management                    & 2131               & 6147             \\
$C_{23}$ & Misc. IV                           & 2835               & 6895             \\
$C_{24}$ & Internet of Things                 & 845                & 4295             \\
$C_{25}$ & Smart Network                      & 802                & 2920             \\
$C_{26}$ & Power System                       & 809                & 2115             \\
$C_{27}$ & Assessment                         & 2347               & 6676             \\
$C_{28}$ & Learning                           & 761                & 3254             \\
$C_{29}$ & Cyber Security                     & 1242               & 16868            \\
$C_{30}$ & Cyber Attack & 804 & 3280             \\ \bottomrule
\end{tabular}
\caption{Summary of keyword clusters}
\label{tab:cluster}
\end{table}
\clearpage

\begin{landscape}
\section{Categorizations in Other Survey Papers}
\begin{longtable}[c]{@{}llll@{}}
\toprule
\textbf{ID} & \textbf{Paper} & \textbf{Reference clusters} & \textbf{CyLit clusters} \\* \midrule
\endhead
\bottomrule
\endfoot
\endlastfoot
1 & \citet{cordonsky_deeporigin_2018} & \begin{tabular}[c]{@{}l@{}}deep learning, \\ malware classification\end{tabular} & \begin{tabular}[c]{@{}l@{}}cyber terminology, security breach, \\ malware, misc, data management,   \\ learning, cyber security\end{tabular} \\ 
\hline 2 & \citet{yousefi-azar_autoencoder-based_2017} & \begin{tabular}[c]{@{}l@{}}deep learning, \\ intrusion detection, \\ malware detection\end{tabular} & \begin{tabular}[c]{@{}l@{}}detection, cyber system management, \\ technology management, misc,  \\ electronic control, malware,\\ data management, learning\end{tabular} \\
\hline 3 & \citet{nguyen_cyberattack_2018} & \begin{tabular}[c]{@{}l@{}}deep learning, \\ intrusion detection\end{tabular} & \begin{tabular}[c]{@{}l@{}}detection, mobile, cyberphysical devices, \\ learning, cyber security\end{tabular} \\
\hline 4 & \citet{alrawashdeh_toward_2016} & \begin{tabular}[c]{@{}l@{}}deep learning, \\ intrusion detection\end{tabular} & \begin{tabular}[c]{@{}l@{}}detection, computational intelligence, \\ misc, cyber security, cyber attack\end{tabular} \\
\hline 5 & \citet{alom_network_2017} & \begin{tabular}[c]{@{}l@{}}deep learning, \\ intrusion detection\end{tabular} & \begin{tabular}[c]{@{}l@{}}detection, cyber terminology, learning,\\ cyber regulations \& public policy,   \\ data management, system security, \\ security breach, malware, misc\end{tabular} \\
\hline 6 & \citet{abdulhammed_deep_2019} & \begin{tabular}[c]{@{}l@{}}deep learning, \\ intrusion detection\end{tabular} & \begin{tabular}[c]{@{}l@{}}detection, cyber terminology, misc,\\ computational intelligence, security breach\end{tabular} \\
\hline 7 & \citet{cox_signal_2015} & \begin{tabular}[c]{@{}l@{}}deep learning, \\ file type identification\end{tabular} & \begin{tabular}[c]{@{}l@{}}computational intelligence, misc, \\ assessment, learning, cyber security\end{tabular} \\
\hline 8 & \citet{fernandez_maimo_self-adaptive_2018} & \begin{tabular}[c]{@{}l@{}}deep learning, \\ intrusion detection\end{tabular} & \begin{tabular}[c]{@{}l@{}}detection, cyber terminology, \\ misc, assessment, learning\end{tabular} \\
\hline 9 & \citet{kessler_information_2020} & \begin{tabular}[c]{@{}l@{}}healthcare, \\ actions of people\end{tabular} & \begin{tabular}[c]{@{}l@{}}cyber system management, healthcare, \\ cyber security, system security, \\ technology management\end{tabular} \\
10 & \citet{colias_infotech_2004} & \begin{tabular}[c]{@{}l@{}}healthcare, \\ actions of people\end{tabular} & \begin{tabular}[c]{@{}l@{}}cyber system management, healthcare, \\ misc, malware, technology management, \\ cyberphysical devices,  electronic control, \\ cyber crimes, data   management\end{tabular} \\
\hline 11 & \citet{priestman_phishing_2019} & \begin{tabular}[c]{@{}l@{}}healthcare, \\ actions of people\end{tabular} & \begin{tabular}[c]{@{}l@{}}cyber system management, security breach, \\ cyber attack, internet of   things\end{tabular} \\
\hline 12 & \citet{dameff_clinical_2019} & \begin{tabular}[c]{@{}l@{}}healthcare, \\ systems and technology failures\end{tabular} & cyber security, healthcare, misc \\
\hline 13 & \citet{kim_risk_2020} & \begin{tabular}[c]{@{}l@{}}healthcare, \\ systems and technology failures\end{tabular} & smart network, healthcare, system security \\
\hline 14 & \citet{moshi_evaluation_2019} & \begin{tabular}[c]{@{}l@{}}healthcare, \\ systems and technology failures\end{tabular} & healthcare, technology management, mobile \\
\hline 15 & \citet{leong_cyber_2020} & \begin{tabular}[c]{@{}l@{}}healthcare, \\ failed internal process\end{tabular} & \begin{tabular}[c]{@{}l@{}}cyber crimes, healthcare, \\ finance \& economics, \\ internet of things\end{tabular} \\
\hline 16 & \citet{akinsanya_towards_2020} & \begin{tabular}[c]{@{}l@{}}healthcare, \\ failed internal process\end{tabular} & \begin{tabular}[c]{@{}l@{}}assessment, cyber security, \\ healthcare, system security\end{tabular} \\
\hline 17 & \citet{williams_cybersecurity_2015} & \begin{tabular}[c]{@{}l@{}}healthcare, \\ failed internal process\end{tabular} & \begin{tabular}[c]{@{}l@{}}healthcare, mobile, system security, \\ security breach, system resilience, \\ cyber security\end{tabular} \\
\hline 18 & \citet{aditya_riskwriter_2018} & \begin{tabular}[c]{@{}l@{}}cyber insurance, \\ organization eligibility\end{tabular} & \begin{tabular}[c]{@{}l@{}}finance \& economics, misc, \\ data management, system security\end{tabular} \\
\hline 19 & \citet{bartolini_using_2019} & \begin{tabular}[c]{@{}l@{}}cyber insurance,\\ organization eligibility\end{tabular} & cyber security, system security \\
\hline 20 & \citet{tondel_differentiating_2016} & \begin{tabular}[c]{@{}l@{}}cyber insurance, \\ organization eligibility\end{tabular} & assessment, cyber security \\
\hline 21 & \citet{yang_incentive_2019} & \begin{tabular}[c]{@{}l@{}}cyber insurance, \\ organization eligibility\end{tabular} & \begin{tabular}[c]{@{}l@{}}cyber system management, \\ assessment, cyber security, misc\end{tabular} \\
22 & \citet{bushnell_cyber-warranties_2018} & \begin{tabular}[c]{@{}l@{}}cyber insurance, \\ contract design\end{tabular} & \begin{tabular}[c]{@{}l@{}}system security, cyberphysical devices, \\ system resilience, assessment,   \\ cyber security\end{tabular} \\
\hline 23 & \citet{knight_framework_2020} & \begin{tabular}[c]{@{}l@{}}cyber insurance, \\ contract design\end{tabular} & \begin{tabular}[c]{@{}l@{}}detection, system security, \\ finance \& economics, \\ malware, cyber   crimes, \\ system resilience, cyber security\end{tabular} \\
\hline 24 & \citet{eling_what_2019} & \begin{tabular}[c]{@{}l@{}}cyber insurance, \\ contract design\end{tabular} & cyber crimes, assessment, misc \\
\hline 25 & \citet{dou_insurance_2020} & \begin{tabular}[c]{@{}l@{}}cyber insurance, \\ contract design\end{tabular} & \begin{tabular}[c]{@{}l@{}}detection, assessment, cyber security, \\ system security\end{tabular} \\
\hline 26 & \citet{laszka_cyber-insurance_2018} & \begin{tabular}[c]{@{}l@{}}cyber insurance, \\ insured self-reporting\end{tabular} & \begin{tabular}[c]{@{}l@{}}finance \& economics, learning, \\ cyber security, misc\end{tabular} \\
\hline 27 & \citet{panda_post-incident_2019} & \begin{tabular}[c]{@{}l@{}}cyber insurance, \\ insured self-reporting\end{tabular} & \begin{tabular}[c]{@{}l@{}}system security, technology management, \\ system resilience, algorithm, cyber security\end{tabular} \\
\hline 28 & \citet{tonn_cyber_2019} & \begin{tabular}[c]{@{}l@{}}cyber insurance, \\ cyber insurance awareness\end{tabular} & cyber system management, cyber security \\
\hline 29 & \citet{pavel_cyber_2020} & \begin{tabular}[c]{@{}l@{}}cyber insurance, \\ cyber insurance awareness\end{tabular} & \begin{tabular}[c]{@{}l@{}}cyber regulations \& public policy, \\ cyber security, misc\end{tabular} \\
\hline 30 & \citet{lau_cybersecurity_2020} & \begin{tabular}[c]{@{}l@{}}cyber insurance, \\ cyber insurance awareness\end{tabular} & smart network, cyber security, algorithm \\
\hline 31 & \citet{farao_secondo_2020} & \begin{tabular}[c]{@{}l@{}}cyber insurance, \\ cost-benefit aspect\end{tabular} & \begin{tabular}[c]{@{}l@{}}cyber system management, assessment,\\ finance \& economics, misc, \\ security breach, cyberphysical devices, \\ electronic control, cyber security\end{tabular} \\
\hline 32 & \citet{tosh_three_2017} & \begin{tabular}[c]{@{}l@{}}cyber insurance,\\ cost-benefit aspect\end{tabular} & cyber security, algorithm, misc \\
\hline 33 & \citet{young_framework_2016} & \begin{tabular}[c]{@{}l@{}}cyber insurance, \\ cost-benefit aspect\end{tabular} & \begin{tabular}[c]{@{}l@{}}finance \& economics, \\ cyber system management, \\ cyber security\end{tabular} \\
34 & \citet{martinelli_preventing_2018} & \begin{tabular}[c]{@{}l@{}}cyber insurance, \\ cost-benefit aspect\end{tabular} & \begin{tabular}[c]{@{}l@{}}cyber system management, \\ \\ system security, finance \& economics, \\ cyber security, cyberphysical devices, \\ misc, system resilience, cyber terminology\end{tabular} \\
\hline 35 & \citet{martinelli_optimal_2018} & \begin{tabular}[c]{@{}l@{}}cyber insurance, \\ cost-benefit aspect\end{tabular} & \begin{tabular}[c]{@{}l@{}}system security, security breach, \\ system resilience, assessment, \\ cyber security\end{tabular} \\
\hline 36 & \citet{kshetri_evolution_2020} & \begin{tabular}[c]{@{}l@{}}cyber insurance, \\ cost-benefit aspect\end{tabular} & \begin{tabular}[c]{@{}l@{}}cyber system management, \\ technology management, system resilience, \\ assessment, cyber security\end{tabular} \\* \bottomrule
\caption{Selected $36$ papers mentioned and labeled in other survey papers}
\label{tab:papers_in_survey}\\
\end{longtable}
\end{landscape}

\clearpage

\section{Experiments with ChatGPT}
\label{appendix_sec:chatgpt_experiments}
We conduct a series of experiments to assess the capability of LLM, specifically ChatGPT-4,  on literature reviews. As documented in Table \ref{tab:chatgpt-experiment}, these experiments were designed to explore various applications of ChatGPT in literature review tasks.

In the experiments, we focus on categorizing 36 selected papers on cyber risk. This experiment was replicated three times to evaluate the consistency of ChatGPT's outputs. Experiments II and III built upon the first by introducing more guided information: Experiment II involves providing ChatGPT with predefined topics based on human judgment, while Experiment III guides the model to follow methodologies similar to those used in our system. Subsequent experiments, IV through VI, were aimed at assessing ChatGPT's ability to summarize and critically review individual papers. These papers, representing diverse aspects of cyber risk, are sampled from the 36 papers. 

It is important to note that multiple trials were conducted for each experiment and instances of failures in generating outputs were not uncommon. Only those instances where ChatGPT successfully generated results are presented and analyzed. For a detailed examination of the prompts provided to ChatGPT and its responses across these experiments, please see our GitHub repository\footnote{\url{https://github.com/changyuehu/CyLit}}, where the complete chat history is accessible.

\begin{table}[!ht]
\resizebox{\columnwidth}{!}{%
\begin{tabular}{@{}lllll@{}}
\toprule
ID &
  Objective &
  Name &
   &
   \\ \midrule
\multicolumn{1}{l}{1} &
  \multicolumn{1}{l}{\multirow{3}{*}{Literature categorization}} &
  \begin{tabular}[c]{@{}l@{}}Experiment I \\ Cyber risk papers categorization (1)\end{tabular} &
   &
   \\ \cmidrule(l){3-5} 
\multicolumn{1}{l}{2} &
  \multicolumn{1}{l}{} &
  \begin{tabular}[c]{@{}l@{}}Experiment I\\ Cyber risk papers categorization (2)\end{tabular} &
   &
   \\  \cmidrule(l){3-5} 
\multicolumn{1}{l}{3} &
  \multicolumn{1}{l}{} &
  \begin{tabular}[c]{@{}l@{}}Experiment I \\ Cyber risk papers categorization (3)\end{tabular} &
   &
   \\ \midrule
\multicolumn{1}{l}{4} &
  \multicolumn{1}{l}{\multirow{2}{*}{\begin{tabular}[c]{@{}l@{}}Literature categorization \\ with additional information\end{tabular}}} &
  \begin{tabular}[c]{@{}l@{}}Experiment II \\ Cyber risk papers categorization with given topics\end{tabular} &
   &
   \\ \cmidrule(l){3-5} 
\multicolumn{1}{l}{5} &
  \multicolumn{1}{l}{} &
  \begin{tabular}[c]{@{}l@{}}Experiment III\\  Cyber risk papers categorization with given methods\end{tabular} &
   &
   \\ \midrule
\multicolumn{1}{l}{6} &
  \multicolumn{1}{l}{\multirow{3}{*}{Review of individual papers}} &
  \begin{tabular}[c]{@{}l@{}}Experiment IV \\ Cyber risk paper review: Yousefi-Azar et al. (2017)\end{tabular} &
   &
   \\ \cmidrule(l){3-5} 
\multicolumn{1}{l}{7} &
  \multicolumn{1}{l}{} &
  \begin{tabular}[c]{@{}l@{}}Experiment V \\ Cyber risk paper review: Kessler et al. (2020)\end{tabular} &
   &
   \\ \cmidrule(l){3-5} 
\multicolumn{1}{l}{8} &
  \multicolumn{1}{l}{} &
  \begin{tabular}[c]{@{}l@{}}Experiment VI \\ Cyber risk paper review: Nguyen et al. (2018)\end{tabular} &
   &
   \\ \bottomrule
\end{tabular}%
}
\caption{Experiments with ChatGPT}
\label{tab:chatgpt-experiment}
\end{table}

\end{appendices}
\end{document}